\documentclass{emulateapj}

\usepackage{natbib,hyperref,ifthen,psfig,graphicx}
\usepackage{epstopdf}
\usepackage{soul,color}
\usepackage{rotating}
\newcommand{\arcse}{\!\!\hbox{$^{\prime\prime}$}}

\def\mathnew{\mathsurround=0pt}
\def\simov#1#2{\lower 2.5pt\vbox{\baselineskip0pt \lineskip-.5pt
\ialign{$\mathnew#1\hfil##\hfil$\crcr#2\crcr\sim\crcr}}}

\def\simgreat{\mathrel{\mathpalette\simov >}}
\newcommand{\MeV}{Me\kern-0.11em V}
\newcommand{\keV}{ke\kern-0.11em V}

\newcommand{\cha}{{\it Chandra\/}}
\newcommand{\xmm}{{\it XMM-Newton\/}}

\newcommand{\galex}{{\it GALEX \/}}

\newcommand{\hst}{{\it HST\/}}

\newcommand{\ecmss}{erg~cm$^{-2}$ s$^{-1}$}
\newcommand{\es}{erg~s$^{-1}$}
\newcommand{\omegam}{$\Omega_{m,0}$\/}
\newcommand{\omegal}{$\Omega_{\lambda,0}$\/}

\newcommand{\Mbh}{\ensuremath{M_{\bullet}}}
\newcommand{\Rsun}{\ensuremath{{\rm R}_{\odot}}} 
\newcommand{\Rstar}{\ensuremath{R_{\ast}}} 
 
\newcommand{\Mstar}{\ensuremath{M_{\ast}}} 
\newcommand{\Msun}{\ensuremath{{\rm M}_{\odot}}}

\newcommand{\ciao}{{\it CIAO\/}}

\slugcomment{Draft version August 23, 2008, accepted for publication in the Astrophysical Journal}

\shorttitle{A Tidal Disruption Flare in Abell 1689}
\shortauthors{Maksym, Ulmer and Eracleous}

\begin{document}


\title{A Tidal Disruption Flare in Abell 1689 \\
    from an Archival X-ray Survey of Galaxy Clusters}


\author{W. Peter Maksym and M. P. Ulmer}
\affil{Department of Physics and Astronomy, Northwestern University,
    Evanston, IL 60208-2900}
\email{pmaksym@u.northwestern.edu}
\author{Michael Eracleous}
\affil{Department of Astronomy and Astrophysics and Center for Gravitation Wave Physics, \\
The Pennsylvania State University, 525 Davey Lab, University Park, PA 16803\\}




\begin{abstract}

Theory suggests that a star making a close passage by a supermassive black hole at the center of a galaxy
can under most circumstances be expected to emit a giant flare of
radiation as it is disrupted and a portion of the resulting stream of
shock-heated stellar debris falls back onto the black hole itself.  We
examine the first results of an ongoing archival survey of galaxy
clusters using \cha\ and \xmm-selected data, and report a likely
tidal disruption flare from \object{SDSS J131122.15-012345.6} in \object{Abell 1689}.
The flare is observed to vary by a factor of $\gtrsim30$ over at least
2 years, to have maximum
$L_X(0.3-3.0$\ keV$)\gtrsim5\times10^{42}$\ erg\ s$^{-1}$ and to emit as a
blackbody with $kT\sim0.12$\ keV.  From the galaxy population as
determined by existing studies of the cluster, we estimate a tidal
disruption rate of $1.2\times10^{-4}\ $galaxy$^{-1}$\ year$^{-1}$ if
we assume a contribution to the observable rate from galaxies whose range of luminosities
corresponds to a central black hole mass (\Mbh ) between $10^6$ and $10^8$ \Msun.
\end{abstract}





\section{Introduction}

It is now generally accepted that massive black holes (MBHs)
reside at the centers of massive galaxies \citep{Magorrian98}.  These
central massive ($\geq 10^6 M_{\odot}$) black holes are expected to
play an integral role in the formation and evolution of galaxies, but
many uncertainties remain in this picture.

The low-mass cutoff for the central black hole mass spectrum is
unknown and the population is inferred from the empirical relationship
between MBHs and their host spheroid's velocity dispersion and
associated luminosity \citep[e.g.,][]{Magorrian98,Marconi03,Lauer07}
or AGN-based models of growth by accretion \citep{Marconi04}.  While
some dwarf galaxies appear to host MBHs and are classified as
AGNs, the active fraction decreases with decreasing mass and many
common assumptions regarding the black hole distribution with respect
to galaxy mass break down at lower \Mbh\ \citep{GHB08}.  Indeed, an
unknown fraction of dwarf galaxies may host no MBH at all but merely a
nuclear stellar cluster \citep[e.g.][]{FerrareseEtAl06}.


Outside of observations of the Galactic center \citep{Ghez03,
  Genzel03}, direct evidence of central black holes in the form of
accretion events in non-AGN (normal) galaxies has been sparse. The
most direct observational evidence of accretion events takes the form
of flares produced by a star that passes close to the central black
hole \citep{Rees88}. The passage is so close that the star is disrupted by the tidal
forces it experiences.  In this case, $\gtrsim\ 1/2$ of its mass is
expelled into space as the remainder falls back to accrete onto the
black hole \citep{KM96}.  Although much theoretical work has been done to describe the expected evolution of the debris and behavior of the resulting flare from an analytical \citep[][e.g.]{Loeb97,Ulmer99} or numerical approach \citep{EK89,KNP93b,ALP00,BogdanovicEtAl04,SQ09}, on the order of a dozen solid observational examples have been so classified to date.   See \cite{Gezari09} for an summary of observations and \citep{Alexander05} for a review of the theory.

Almost all the observations of tidal flares (or flare candidates) have
been found in surveys of the field \citep{Esquej07,Esquej08, Donley02}
or in in the UV \citep{Gezari09,Gezari06}, and while our work
was already in progress has there been an event reported to take place
in a cluster of galaxies \citep{Cappelluti09}.

Our paper is the first in a planned series of studies aimed
specifically at monitoring clusters of galaxies for flares using
archival \cha\ and \xmm\ data.  This paper is based on observations of
the first cluster we have chosen to study, \object{Abell 1689} (Luminosity distance modulus 39.68, $z=0.183$\footnote{From the NASA/IPAC Extragalactic Database (NED) which is operated by the Jet Propulsion Laboratory, California Institute of Technology, under contract with the National Aeronautics and Space Administration.}) .  A1689 was an
ideal choice for this survey because it exhibits many characteristics
which maximized the chance of observing tidal flares given existing
rate predictions.

\object{Abell 1689} is among the richest of the Abell clusters, one of only 9
that has a richness of 4 or higher (Abell 665 is the only cluster of
richness 5).  At a redshift of $z\sim\ 0.18$, with a virial radius
$r_{200}$ of 1.13 $h^{-1}$ Mpc and hence about 10\arcmin, the entire
core of the cluster fits within a single \cha\ ACIS chip \citep{AM04}.  Its
richness has made it popular for examinations of the Sunyaev-Zeldovich
effect, structure analysis, and lensing studies so it is well-observed
in X-rays, optical and other wavelengths (e.g. Hubble) and hence an
ideal starting point for our survey.  With this extensive coverage
(described in detail below) of a rich cluster of galaxies whose members are all
effectively at the same redshift, we have thus been able to estimate
the rate at which the galaxies experience tidal flare events.

In this paper we present the discovery of a transient event that we
demonstrate to be most probably due to emission from a tidal flare and
whose temperature is arguably the highest yet observed. For,  
although the derived peak observed flare
temperature matches the highest reported peak blackbody temperature
\cite{Cappelluti09}, our result was based on data from an
instrument that provides about $\sim 3 $ times better energy
resolution than was used to derive the temperature of the
\cite{Cappelluti09} flare.  

Furthermore, by monitoring so many
galaxies it the field of view (FOV) at the same time, it was possible
for us to set a limit to the frequency of flare events in a well
defined galaxy population.  Depending on the hypothesized event rate
\citep[e.g.,][]{Wang04,Merritt09}, our results may be used to delimit
either the event rate or the very existence of massive black holes in
dwarf galaxies.  Tighter constraints on the event rate, as derived
from the work presented here, are also interesting for predictions of
event rates of the {\it Light Interferometer Space Antenna}\footnote{http://lisa.nasa.gov/} (hereafter
{\it LISA}) which will be sensitive in the mass range of $\Mbh \lesssim
10^{7}\;\Msun$
\citep{Jennrich04,SHMV04,SHMV05,Sigurdsson03,Kobayashi04}.

Throughout this paper, we adopt concordant cosmological parameters of
$H_0=70\ $km$^{-1}$ sec$^{-1}$ Mpc$^{-1}$, \omegam=0.3 and \omegal=0.7,
and calcuate distances using \cite{Wright06}. All coordinates are
J2000.  The galactic column density of neutral hydrogen for \object{Abell 1689}
is $1.82\times10^{20}\ $cm$^{-2}$, derived using \cite{Dickey90} values from the {\it
  colden} tool in \ciao\ \citep{ciao06} unless otherwise stated.  All X-ray fluxes and luminosities
used in this paper have been corrected for Galactic absorption using
these values.

\section{Observations and Data}

\subsection{Overview}

In order to identify tidal flare candidates, we begin by compiling a
catalog of X-ray point sources and identify the characteristic
properties that distinguish such flares from other X-ray point sources
such as AGNs, supernovae (SNe), Gamma Ray Bursts (GRBs) and flaring M-dwarf
stars.  We can expect most variable X-ray sources to be AGNs.  Unlike
AGNs which typically have a hard power-law spectrum
\citep[e.g.][]{Beckmann06}, tidal flares will have extremely soft ($kT\lesssim\ $0.1~keV) blackbody spectra.  Tidal flares should have much
greater variability than AGNs near their peak, but that timescale may
be of short ($\lesssim 1\ $yr) duration.  A demonstrated pre-flare quiescent
period as characterized by a lack of significant X-ray emission would
also distinguish a tidal flare from an AGN.  Redshifted background AGNs, supernovae, GRBs and M-dwarfs may also have soft X-ray emission, but background AGNs may be identified by
their optical spectra and M-dwarfs will also produce characteristic optical emission.  
 
X-ray-luminous SNe present a subtle contamination challenge to an archival survey given the likely impracticality of optical follow-up on long timescales and possible lack of sufficient light curve coverage to describe a decay index.  But SNe may be eliminated as an explanation through a combination of the flare's peak luminosity and observed rate of decay. 
SNe and GRBs typically produce X-rays on a timescale of days or months and may be be identified optically as well.  If the strongest modes of soft, luminous X-ray emission arise first from the prompt shock breakout from the supernova progenitor's surface \citep[e.g.][]{Soderberg08}, then from the expansion of the ejecta as they interact with the surrounding medium \citep{SP06,Immler08}, SNe may be correspondingly distinguished from tidal flares.  

The most X-ray-luminous SNe due to ejecta expansion \citep[e.g.][]{Immler03,SP06,Immler08} may have long ($\gtrsim$ years) periods of shallow decay.  But whether described by thermal or power law spectra, they have not been demonstrated to have peak luminosities significantly in excess of $\sim10^{42}$ \es\ and typically do not exceed $\sim10^{40}$ \es.  The progenitor breakout phase may have a supersoft thermal X-ray luminosity in excess of $10^{43}$ \es, as is also expected from a tidal flare, but this breakout has also shown a fast rise and exponential decay on a timescale of $\sim100$ sec \citep{Soderberg08}.  Lack of strong variability over timescales of typical \cha\ or \xmm\ galaxy cluster observations (tens of kiloseconds) should thus exclude breakout from the progenitor as an explanation for a supersoft X-ray flare of comparable luminosity.
 
The location of
the event relative to the host galaxy can be used to rule out a tidal
flare event if the location is not consistent with the core of a host
galaxy, such as for a supernova outside the galaxy core or a
background GRB with no host.  As such, we have sought to associate
flare candidates with host galaxies using optical data, and to
constrain the Rayleigh-Jeans tail of any candidate events with UV data where
available.

For our primary dataset, we obtained \cha\ ACIS archival data for six
epochs spanning a period of 8 years, from April, 2000 to March, 2007.
In addition, we examined \xmm\ data from December, 2001.  X-ray epochs
are indicated in Table \ref{obs-1}.  In order to follow up specific
sources, we supplemented our dataset with archival \hst\ WFPC2 images
with filters F606W and F814W.  For our candidate event, we obtained a
March 2009 spectrum of the putative host galaxy with the
$Hobby-Eberly\ Telescope$ (HET).

\subsection{Chandra X-ray Observations}  

\subsubsection{Data Reduction and Candidate Selection}

We reprocessed the ACIS event files using \ciao\ to remove pixel
randomization.  According to standard \ciao\ data processing threads,
we also applied TGAIN and CTI corrections, and VFAINT mode cleaning
where necessary.  We filtered the events to use grades 0, 2, 3, 4 and
6 and energies between 0.3 and 10 \keV.  During all observations, the
target was placed at the ACIS-I aimpoint. All observations used
ACIS chips I0-4 and S2, except for obsid 7701 which used the
ACIS-I chips only.  ACIS suffers from charge readout issues which may be remedied through use of a {\it destreak} software tool, but the problem is generally only significant on the S4 chip which was inactive in all epochs.  We thus did not apply {\it destreak}.
To account for background cosmic-ray induced flares, we filtered
individual chips for 3$\sigma$ deviations using the {\it CXC\/}
analyze\_ltcrv.sl script after excluding bright sources found using
{\it wavdetect\/}.  Minor flaring, of duration 20 ksec or longer, was
present in all epochs; data from the corresponding time intervals were
excluded from further processing.

We defined several bands in accordance with the {\it ChaMP\/}
definitions found in \cite{ChaMPCat}: B (0.3--8~keV), S
(0.3--2.5~keV), H (2.5--8~keV), S1 (0.3--0.9~keV), and S2
(0.9--2.5~keV).  In each band, we produced weighted exposure maps and
compiled a source list with {\it wavdetect\/}.  {\it wavdetect\/} used
a $\sqrt{2}$ scale progression of $1, \sqrt{2}, 2, 2\sqrt{2}, 4,
4\sqrt{2}, 8$ on band images binned by a factor of two, covering a
square of width 4096 pixels in physical coordinates that
enclosed the whole of the ACIS-I array and hence the cluster core.  We
used a significance threshold of $10^{-6}$ for {\it wavdetect\/},
corresponding to an approximate maximum of one false source per chip
when singly binned, and $\sim 1$ false source over the whole array
when binned by two.

We also created merged, fluxed mosaic images in each band with
1\arcsec\ pixels, which we broke into overlapping images with
dimensions of $1024\times 1024$ pixels for the purposes of running
wavdetect.

Once we had compiled a master source list, we eliminated conspicuous
edge sources and consolidated redundant multiple detections of single sources across epochs, retaining the astrometric positions that demonstrated the best accuracy.  We also eliminated most sources within 0\farcm5 ($\sim 95$ kpc) of the
cluster center, determining them to be unreliable based on {\it MARX}
simulations using a $\beta$ model for the Intergalactic
Medium.  We had found these sources to be indistinguishable from Poisson fluctuations in the intracluster diffuse emission.  We then visually selected overlapping sources for
consolidation, using the $95\%$ encircled energy radii as a guideline
and giving preference to positions of detections with larger SNR, or
to \cha\ sources in case of additional detection by \xmm.

For initial X-ray source classification, we determined count rates
using the \cite{Feigelson02} polynomial approximations for the 95\%
encircled energy radius of each epoch, $r_{95}$ and a background
annulus between the 99\% encircled energy radius $r_{99}$ and $5\times
r_{99}$.  We used {\it dmextract} to determine count rates for each
source and determined a subset of sources that across epochs
demonstrated (0.3-2.5 keV) variability by a factor of a few, $\gtrsim\ 4-5$.  All
errors use the \cite{Gehrels86} approximation for Poisson
statistics.

Within the set of sources that demonstrated significant variability,
we examined basic spectral properties to determine likely soft
blackbodies and rule out AGNs at low-to-moderate redshifts, which we
expect to be the most common point source at galactic latitude
$b\gtrsim\ 60$.  We define hardness ratios between bands
$HR_{1,2}=(c_2-c_1)/(c_2+c_1)$ where $c_n$ is the net number of counts
in {\it ChaMP}-defined band $n$ (minimum $c_n=0$).  Given a typical AGN
continuum with X-ray photon index $\Gamma \lesssim 2.5$ ($F_X\propto {\rm E}^{-\Gamma}$), $HR_{H,S2}$ will be
sensitive to unusually soft spectra except in the presence of a
significant hard background.  Even at $L_X$(0.3--8 keV) $\gg 10^{41}$ \es\ , the
exponential cutoff on the Wien side of a soft blackbody spectrum will
result in few events in the H band, if any.  But $HR_{S2,S1}$ will
be an exceptional diagnostic for blackbodies in the temperature range
anticipated for tidal disruptions if peak emission is somewhere in the
vicinity of the \cha\ bandpass cutoff at $kT \lesssim 0.3\ $keV.

We also compared variable sources to the Sloan Digital Sky Survey (SDSS) catalog, Data Release 6 \citep{SDSSdr6}, eliminating candidates that corresponded to a photometric redshift consistent with background objects ($z\gg0.2$), most of which were also flagged by SDSS as likely QSOs.  We also eliminated stars, determined to be bright non-QSO objects designated as stars by SDSS. 

Of the 226 distinct sources we examined, almost all varied in count rate in one or more X-ray bands across epochs, but such variation was usually statistically insignificant.  For all but four sources, counting statistics in the hard (2.5--8 keV) band were insufficient to conclusively demonstrate variability.  Among the highly variable sources, i.e. those which varied by $\times4$ or more in B (0.3--8 keV), only a few demonstrated an unambiguous early X-ray {\it nondetection} (such that the nondetection could not be attributed to the choice of the background extraction region 
and local background fluctuations), followed by single follow-on rise in photon flux, as well as 
later flux evolution consistent with steady decline across epochs.  Of this subset of sources, 
only one (source 141) found at $(\alpha, \delta)=13^{h}11^{m}22^{s}, -1\degr23\arcmin45\arcsec$ (J2000) both demonstrated an unusually soft peak spectrum and was not eliminated via SDSS as a background object, a likely QSO or a foreground star. 
All other highly variable sources were also associated with AGNs/QSOs or foreground stars, and also had harder spectra ($HR_{H,S}\geq-1.00$).  Except for sources 78 and 166, which are associated with SDSS stars that have NOMAD \citep{NOMAD05} proper motions $\ga 14$ mas yr$^{-1}$, and source 153 whose variability is inconsistent with a flare's rise and fall, all other sources have $HR_{H,S}\geq -0.64$.

\begin{figure}
\plotone{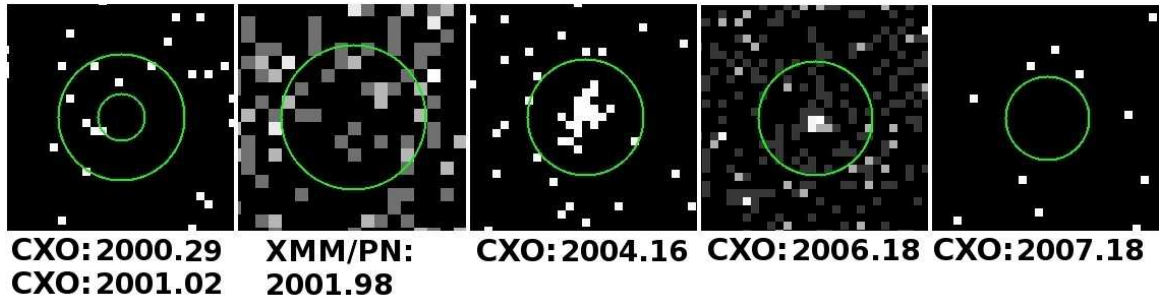}
\caption{Images of the event at the times of observation epochs,
  starting from pre-flare.  Instruments used are indicated: CXO for \cha\ {\it X-ray Observatory} ACIS, and XMM/PN for the \xmm\ PN.  Numbers correspond to the date of observation.  Circles
  describe a 95\% encircled energy radius for \cha\ ACIS or 50\ in the case
  of \xmm\ PN.  Color scales are uniform and scaled to provide a good
  contrast at epoch 2006.18.  Note that this saturates the color scale
  at epoch 2004.16, the observation nearest to peak.  \cha\ images are
  binned from (0.3-10 keV) event files by a factor of two, \xmm\ unfiltered by energy and binned by a factor of 32.
  Pre-flare \cha\ epochs have been merged with extraction circles from
  both epochs indicated (as per differing off-axis angles).
 }
\label{lineup}
\end{figure}

The X-ray images of the source observed across all epochs and its immediate vicinity are displayed in Figure \ref{lineup}.  

For comparison, Table \ref{sigvar} shows a subset containing the 22 most significantly variable sources, such that all had count rates which varied by both at least $2\sigma$ and by $\times4$ over at least 2 epochs of \cha\ coverage.  Due to variations in aimpoint, roll angle and choice of active ACIS chipset across epochs, many sources fell within the \cha\ FOV for a fraction of the available observations, as indicated in Table \ref{sigvar} .  
Rough broadband (B) count rates and hardness ratios ($HR\equiv HR_{H,S}$) were derived using
{\it dmextract} with a constant extraction radius of 10\arcsec, sufficient to enclose $\gg 90\%$ enclosed energy even
at large aimpoint offsets, and a concentric background annulus with twice the area of the extraction radius.  Note that for small $\theta$ this will
introduce additional rate uncertainty, due to the modest size of the background annulus and due to the enclosure of additional background within a larger source
extraction radius.  But it also improves the tolerance for positional uncertainty which may be an issue for fainter sources.

\begin{figure}
\plotone{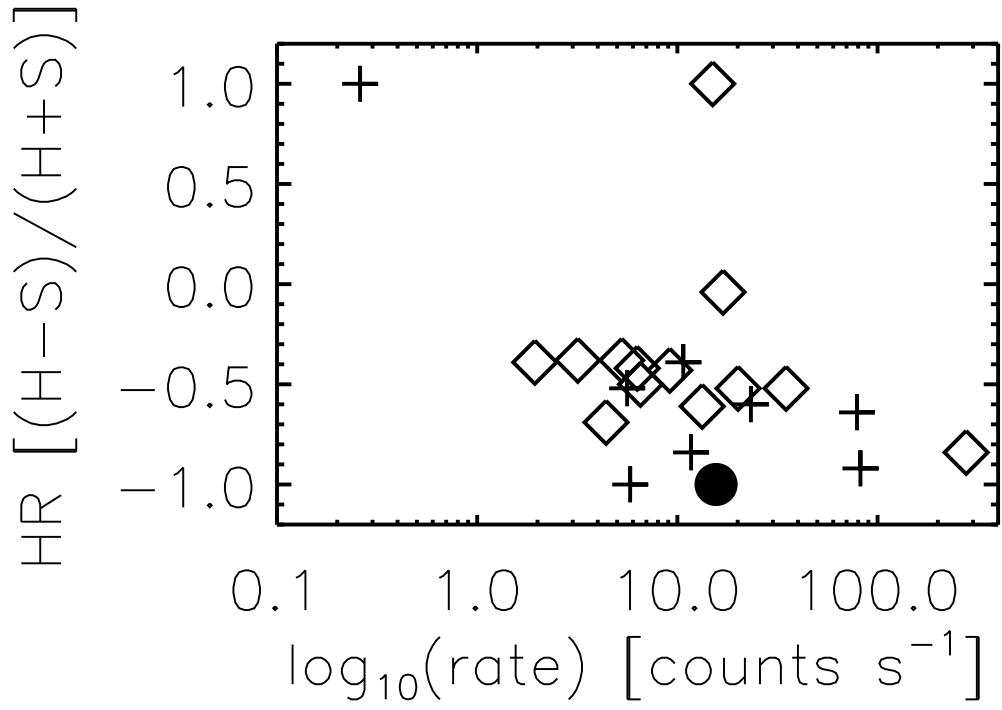}
\caption{Sources from Table. \ref{sigvar} with hardness ratios plotted against corresponding peak count rates.  The filled circle indicates our tidal flare candidate, source 141.  Crosses mark those X-ray sources which correspond to SDSS objects designated as stars.  Diamonds indicate sources all other sources.}
\label{cr_vs_hr}
\end{figure}

In Figure \ref{cr_vs_hr}, we have plotted $HR_{H,S}$ vs. peak count rates from Table \ref{sigvar}.  Compared against the subset of significantly variable sources not corresponding to stars, source 141 is a clear soft outlier.  Most of these non-stellar sources cluster at $HR_{H,S}\sim -0.50$, but source 141 is the lowest.

In very few cases is it possible to place strong limits on the pre-peak absence of an X-ray source.  Peak emission in the first epoch of
coverage accounts for 8 of these objects.  In most cases where the observed peak is later than the first epoch of coverage, manual examination of the event image provides at least some evidence of pre-peak source emission (except for the three objects noted in Table \ref{sigvar}).  
 With those sources eliminated from consideration, all remaining significantly variable sources could not have had steady emission stronger than the $1\sigma$ detection level of $1.9\times 10^{-4}$ ct s$^{-1}$, determined by a single 10 ks observation.

Most objects have one corresponding SDSS object within the \cha\ PSF (a few \arcsec).  But almost all of these objects have been flagged due to traits that are problematic to interpretation as a tidal disruption.  SDSS has flagged 12 optical matches as likely stars due to object
morphology and psf modeling, of which 4 are also flagged as potential QSOs due to colors atypical of stars.  If these flags are accurate descriptions of the objects in question, they would tend to exclude tidal disruptions in normal galaxies
as an explanation for the X-ray variability.  Exceptions to this restriction can be imagined, such as if the SDSS observation were near a flare's X-ray peak and captured the blue Rayleigh-Jeans tail in the case of an object flagged as a possible QSO.  Bright emission from the center of a relatively faint galaxy might also mimic a stellar radial profile.   But without strong evidence from the X-ray emission, the simpler explanation of a star or
QSO is preferred.  Matches can also be found between 4 of these star-like sources with SDSS correspondences and the 1RXH \citep{1RXH} and 2RXP \citep{2RXP} catalogs, suggesting persistent pre-existing sources rather than new flares.

Excluding all 12 of these flagged objects from Table \ref{sigvar} leaves 10 sources, of which only 3 significantly variable objects are classified as SDSS galaxies.  SDSS photometric redshifts indicate source 167 is a likely foreground object, source 157 is a likely
background object, and 141 is a plausible cluster member.  The 7 other unflagged objects have no corresponding SDSS objects and therefore two possible interpretations.  Either they must be background objects which we would not examine as flares for the purpose of this paper (and for
which the most likely interpretation is an AGN, given the composition of the X-ray background, see e.g. \cite{Giacconi01,Szokoly04} and references therein), or as dwarf galaxies below the optical detection threshold, for which a tidal flare would remain a
possibility of interest.  Two such sources, 134 and 153, appear to have an optical counterpart unregistered by SDSS due to proximity to an image flaw.  But for the unmatched sources, the X-ray
lightcurves are either too faint or too sparsely sampled to make strong claims as to whether they are the result of tidal disruptions.  Including those tagged as possible QSOs, there are 7 star-like SDSS-matched sources which do not have
X-ray variability demonstrably inconsistent with a tidal flare.  In \S4 we will consider those as possible contributors to the flare rate, as well as the 7 unmatched objects, but the available data do not
warrant more detailed examination.

The exact behavior of tidal fallback emission remains uncertain and may be expected to harden according to $T_{bb}\propto\Mbh ^{-1/4}$ \citep[e.g.][]{Ulmer99}.  But $HR$ indicates an interestingly soft spectrum from the sole likely A1689 cluster member that is both definitely an
SDSS galaxy and significantly X-ray variable.  According to $HR$, it is one of the two major soft outliers in the significant variability population (the other of which, source 153, demonstrates erratic light curve behavior clearly inconsistent with the simple rise and fall of a tidal flare).  

Having excluded the other X-ray sources from consideration as tidal flare candidates, we will examine source 141 in detail in the remainder of \S2 and \S3.

\subsubsection{Photometry of Source 141}

\paragraph{High State:}

The source had been observed in obsid 5004 on February 28, 2004.  It
had been found by $wavdetect$ in this \cha\ epoch only, and there in
all bands except H.  Within $r_{99}$, \cha\ detected a total of 41
counts in the X-ray B band over 19.9 ksec, with an exceptionally
soft$\ HR_{H,S2}=-1.00$ and $HR_{S2,S1}=-0.49$.  Within $r_{95}$ this
becomes 37 total counts and $HR_{S2,S1}=-0.41$.  With an estimated
$r_{95}$ background contribution of 5.45 counts, this is a
$\sim9.0\sigma$ detection above background fluctuations with a net
rate of $1.6\pm0.3\times 10^{-3}$~s$^{-1}$.  Given the H band source
emission is indistinguishable from background, this becomes
$\sim11.4\sigma$ when restricted to the S band only.

At 3\farcm7 from the cluster core, contributions from the diffuse
ICM and faint galactic haloes are sufficiently significant that care
must be taken in choosing a background extraction region.  Along the
radial direction from the cluster core the background varies by a
factor of $\gtrsim2$ within 20\arcsec\ of the source for an X-ray B
band image of all epochs merged.  But the source is sufficiently
bright in the 2004 epoch that choosing a background annulus within
$5\times r_{99}$ does not significantly affect source estimates.

\paragraph{Pre-flare:}With no detections of the source in two previous 
\cha\ epochs of only 10.5~ksec apiece, we coadded results from obsids
540 and 1663 using epoch-appropriate extractions to determine more
stringent upper bounds for a source that is assumed not to be
significantly variable.  Respectively, the $r_{99}$ count rates are
within $0.5\sigma$ or less than those of the background extraction
annulus.  There are no X-ray B band counts within $r_{50}$ for either
epoch.  Over both pre-flare epochs we find a count rate of
$9.5\times10^{-5}$~s$^{-1}$ within $r_{95}$, still less than the
estimated $r_{95}$ background rate of $2.5\times10^{-4} $~s$^{-1}$.
We therefore estimate a $1\sigma$ upper bound of $1.3\times10^{-4}
$~s$^{-1}$ in excess of the background.

\paragraph{Decay:}A much fainter source was also found by $wavdetect$ 
at the same coordinates after the 2004 outburst, but only in the S and
S1 bands of obsid 7289 beginning March 9th, 2006.  Over 76 ksec there
are 51 total and 11.33 net X-ray B band counts within $r_{99}$.  Using
smaller extraction radii results in the loss of a small fraction of
the source counts but also reduces contributions due to background,
which generally increases with increasing radii due to an increased
fraction of the brighter core regions of the ICM.  Thus, in
determining the significance of the March 2006 emission, which is
faint with respect to the background, we use the 90\% encircled energy
radius enclosed by a background annulus out to $2\times r_{90}$.  From the March 9th, 2006 observations, we find 28 total and 11.48 net X-ray B band counts within 76~ks, i.e.,
$\sim2.2\sigma$ above background.  Although there is no detection from obsid 6930 on March
6th, 2006, we similarly find 18 total and 6.98 net counts within 77~ks, i.e., $\sim1.7\sigma$ above background.  The two March 2006
observations are separated by a time interval equivalent to their
combined duration, i.e., much smaller than the $\sim2$ yr elapsed from
obsid 5004. Because of this short separation, we henceforth treat
obsids 6930 and 7289 as a single 153~ks observation, which leads to a
$\sim3\sigma$ detection of the source with a count rate of
$1.2\pm0.5\times10^{-4}$~s$^{-1}$. 

At only 5 ksec, obsid 7701 produced no detection of the source and no
photons within $r_{95}$ on March 7, 2007.  Taking the emission to be
background, we estimate the $1\sigma$ upper bound of
$3.7\times10^{-4}$~s$^{-1}$ in excess of the background.

\paragraph{Variability:}In order to confirm the degree of variability, 
we used the \ciao\ tool $glvary$, which characterizes a Bayesian
analysis of Poisson statistics with a variability index (VI) based on
an log$_{10}(O)$ where O is the odds ratio \citep{GL92,Rots06}.
Examining both the X-ray S and B bands across all epochs and across
expected rise and fall times, we find VI$\geq 8$ and $O > 5.9$ in all
cases, where VI $\geq 6$ is considered definitely variable and the
maximum VI is 10.  The source shows no significant variability
within its high-state epoch, obsid 5004.

\subsubsection{Least Squares Spectral Fitting}

We fitted the spectra from the high and post-decay states (from
February 2004 and March 2006, respectively ) using {\it XSPEC} v12.4.0
\citep{Arnaud96,Dorman01} with
a variety of blackbody and power law models that include absorption by
line-of-sight neutral hydrogen\footnote{Due to the relatively low redshift of the object, it does not matter if the absorption is assumed to be in our 
galaxy or the
external galaxy. For simplicity we assume the absorption all takes place within our galaxy for the purposes of spectral fitting.} and \cite{chur96} weighting.  We
assumed that the source was a member of the cluster at fixed $z\sim
0.183$, although $z$ did not change by more than a few percent when left
free to vary.  For spectral fitting we used 99\% encircled energy
extraction radii and account for \cha\ ACIS instrumental response by calculating position-dependent response matrices using {\it mkrmf} and {\it mkarf}.

We found the high state emission fitted well a blackbody of temperature of
$kT_{bb}=0.12\pm0.02\ $keV, with $\chi^{2}/\nu=5.89/5.00$.  Fixing
$kT_{bb}$ at this value, we estimate an X-ray flux.  $F_{X}$(0.3--10~keV)$=3.7\pm1.4\times10^{-14}$ \ecmss, with negligible
contribution above 2 keV.  Fitting tended to decrease estimated $N_H$
below the galactic values found by \cite{Dickey90} to unphysically low numbers,
so we froze $N_H$ at the {\it colden} value of
$1.82\times10^{20}\ $cm$^{-2}$.  Corrected for galactic extinction, $F_{X}$(0.3--10~keV)$=4.7\pm1.7\times10^{-14}$ \ecmss.

The fit for a power law produced a photon index $\Gamma=4.86\pm0.48$
with a modestly better fit, $\chi^{2}/\nu=5.36/5.00$ and
uncorrected $F_{X}$(0.3--10~keV)$=5.6\pm2.5\times10^{-14}$ \ecmss, 99.5\% of which is below 2
keV.  Allowing $N_H$ to vary freely increases it to
$1.07\pm0.41\times10^{21}\ $cm$^{-2}$ with $\Gamma=5.8\pm3.5$,
$\chi^{2}/\nu=4.98/5.00$ and uncorrected $F_{X}$(0.3-10~keV)$=5.6\pm2.5\times10^{-14}$ \ecmss.

For the decayed state, blackbody fitting produces $1\sigma$ upper
limits of $N_H=1.6\times10^{22}$~cm$^{-2}$ and $kT_{bb}=0.17$ keV with
$\chi^{2}/\nu=8.35/9.00$, consistent with high state properties, and a
model $kT_{bb}=0.10\pm0.02$, if we fix $N_H$ during the fit at the observed
galactic value.  If we assume no change in spectral shape over 2
years, we estimate an uncorrected $F_{X}$(0.3--10~keV)$=4.1\pm2.8\times10^{-15}$ \ecmss.  Corrected for galactic extinction, $F_{X}$(0.3--10~keV)$=5.1\pm3.5\times10^{-15}$ \ecmss.

A power law fit for the post-decay state with galactic $N_H$ gives
$\Gamma=5.7\pm1.1$, $\chi^{2}/\nu=8.93/10.00$ and
uncorrected $F_{X}$(0.3--10~keV)~$\lesssim1.4\times10^{-14}$ \ecmss.  $\Gamma$ cannot be
reasonably constrained, if $N_H$ is left free to vary.

\begin{figure}
\plotone{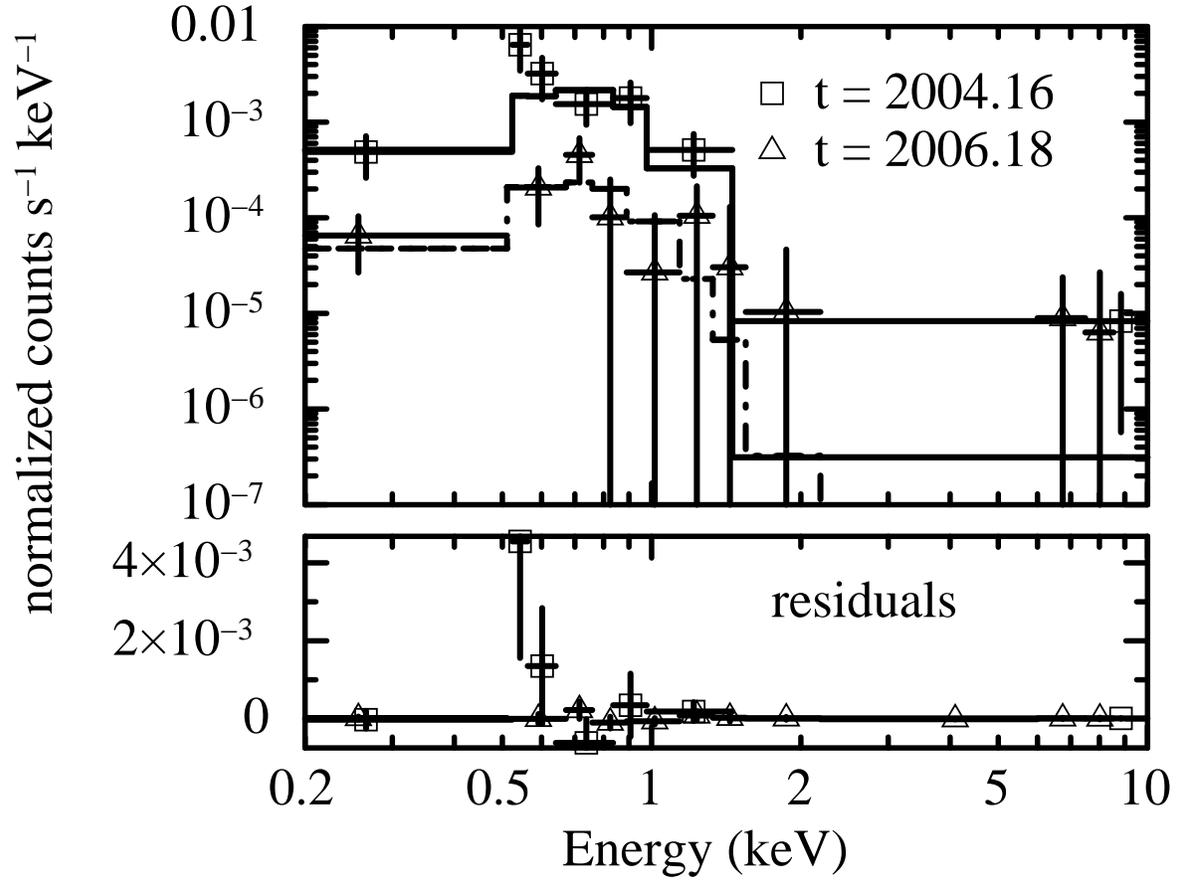}
\caption{Top: fits to the \cha\ spectra using a $kT=0.12$\ keV blackbody
  model at the cluster redshift at epochs 2004.16 (solid line: model, squares: data) and
  2006.18 (dot-dashed line: model, triangles: data).  $1\sigma$ errors are indicated.  Each bin
  signifies 5 photon counts.  Bottom: residuals between model and data, as above.}
\label{xspec}
\end{figure}

An overlaid spectrum of a simultaneous blackbody fit of the high and
low states is presented in Figure \ref{xspec}.  If we assume the combined source flux from pre-flare epochs represents an approximately constant background, we can fit it to high state blackbody temperature, $N_H$ and redshift.  This fit predicts no pre-flare source $F_{X}$(0.3--10~keV) with a $1\sigma$ upper bound
$6.9\times10^{-15}$ \ecmss.  Corrected for galactic extinction, $F_{X}$(0.3--10~keV) $\leq8.6\times10^{-15}$ \ecmss.  If we use $r_{90}$ to reduce likely
background contamination of this estimate, we find an extinction-corrected limit of
$1.6\times10^{-15}$ \ecmss.  With a much shorter observation time than
any epoch, we can constrain the post-flare flux from the March 2007
epoch only very roughly.  Using similar methods to those by which we estimate the
pre-flare flux, even 0 counts within $r_{95}$ lead to a blackbody
extinction-corrected $F_X$(0.3--10~keV)$=8.3\times10^{-15}$ \ecmss\ or a $1\sigma$ upper bound
of $3.5\times10^{-14}$ \ecmss\ for 4 counts within $r_{99}$ in a
single energy bin.

\subsection{Supplementary Observations}

\subsubsection{XMM-Newton Data}

Pipeline processing for \xmm\ found no corresponding X-ray source for
the 40 ksec December 2001 epoch. Our examination of the MOS and PN
event files and the composite broadband image (0.2--12.0 keV) found no
 evidence of a source within \xmm\ $r_{80}$, which at
$\sim20\arcsec$ is significantly larger than the \cha\ positional
uncertainty.  The mean value of the pipeline sensitivity map within $r_{80}$ is
0.0022~s$^{-1}$, equal to the $3\sigma$ upper limit of the encircled photons. 
We fitted a PN spectrum generated using the {\it espec} tool from the \xmm\ SAS software suite, version 7.1.0.  For a $1\sigma$ blackbody with peak $kT_{bb}$, $N_H$ and redshift, we find an uncorrected upper limit $F_X$(0.3--10~keV)$=8.5\times10^{-16}$ \ecmss\ or extinction-corrected $F_X$(0.3--10~keV)$=1.1\times10^{-15}$ \ecmss .  This source region is also included with the \cha\ data in
Figure \ref{lineup}.

We also examined optical and ultraviolet images observed using the \xmm\ Optical Monitor with the filters UVW2 (1800\AA--2300\AA), UVW1 (2300\AA--3700\AA), U (3000\AA--3900\AA), and B (3800\AA--4900\AA).  
The region within $\sim2\arcmin$ of the source is dominated by a ghost
ring in the Optical Monitor, an artifact of increased background due
to reflection of stray light from a bright source.  This contamination
dominates across all filters and prevents extraction of useful optical
or ultraviolet data.

\subsubsection{SDSS and CTIO Data}
  
We examined the Sloan Digital Sky Survey Catalog from Data Release 6
\citep{SDSSdr6} to eliminate flaring red dwarfs as a possible
explanation for variable, extremely soft emission within our field and
found several such objects corresponding to otherwise interesting
X-ray candidate sources.  From our softest sources we found one SDSS
object designated a galaxy by PSF modeling that corresponded to the
Chandra $wavdetect$ position of \object{SDSS J131122.15-012345.6}.  This galaxy
was observed on March 3, 2000 and found to have SDSS $ugriz$
magnitudes of $u=25.27\pm1.14$, $g=21.45\pm0.07$, $r=20.46\pm0.04$,
$i=20.10\pm0.05$ $z=19.63\pm0.11$.  The large $u$ error is likely due
to being a nondetection in the $u$-band, which we confirmed by
examination of $ugriz$ band images.  Note that the characteristic
limiting magnitude for SDSS in the $u$-band is 22.3 so the uncertainty
is likely even higher.

SDSS Photometric redshifts vary by $\sim0.23$ depending upon the method used but suggest the galaxy is likely a member of A1689 given the much smaller
odds of a foreground nonmember and typical unreliability of blind
photometric redshifts at $z\ll 1$.  The SDSS template fitting
algorithm finds a redshift of $z=0.097\pm0.03$ whereas the catalog's
neural network derived redshifts are $z=0.20\pm0.09$ and
$z=0.33\pm0.07$ for the CC2 and ZD1 methods respectively.  We have
also run our own photometric redshift estimates, finding
$z=0.16\pm0.09$ using ANNz neural networking software \citep{ANNz}.

We estimated the redshift with LePhare template fitting software
\citep{LePhare} and its Hubble Deep Field North libraries.  By
incorporating CTIO J- and K-band data from \citet{Stanford02}, where
$M_J=19.12\pm0.12$ and $M_K=17.02\pm0.11$, we find an effective upper
bound of $z=0.20$ and a most likely model of an elliptical galaxy.
The redshift probability distribution is bifurcated with maxima at
$z=0.04$ and 0.20.  The upper maximum reflects an Sa model, which
fits $z=0.19\pm0.04$ if the elliptical model is eliminated from
consideration, and a bolometric absolute magnitude $M_{bol}= -20.17$
at that redshift.  Redshift $z=0.04$, on the other hand, would imply
$M_{bol}=-16.29$.  The fit cannot exclude the possibility of a QSO at
$z\simeq3.44$ which assumes Ly$\alpha$ emission that is undetected in
the gap between the $g$ and $r$ bands.

\subsubsection{HST-WFPC2}

\begin{figure}
\plotone{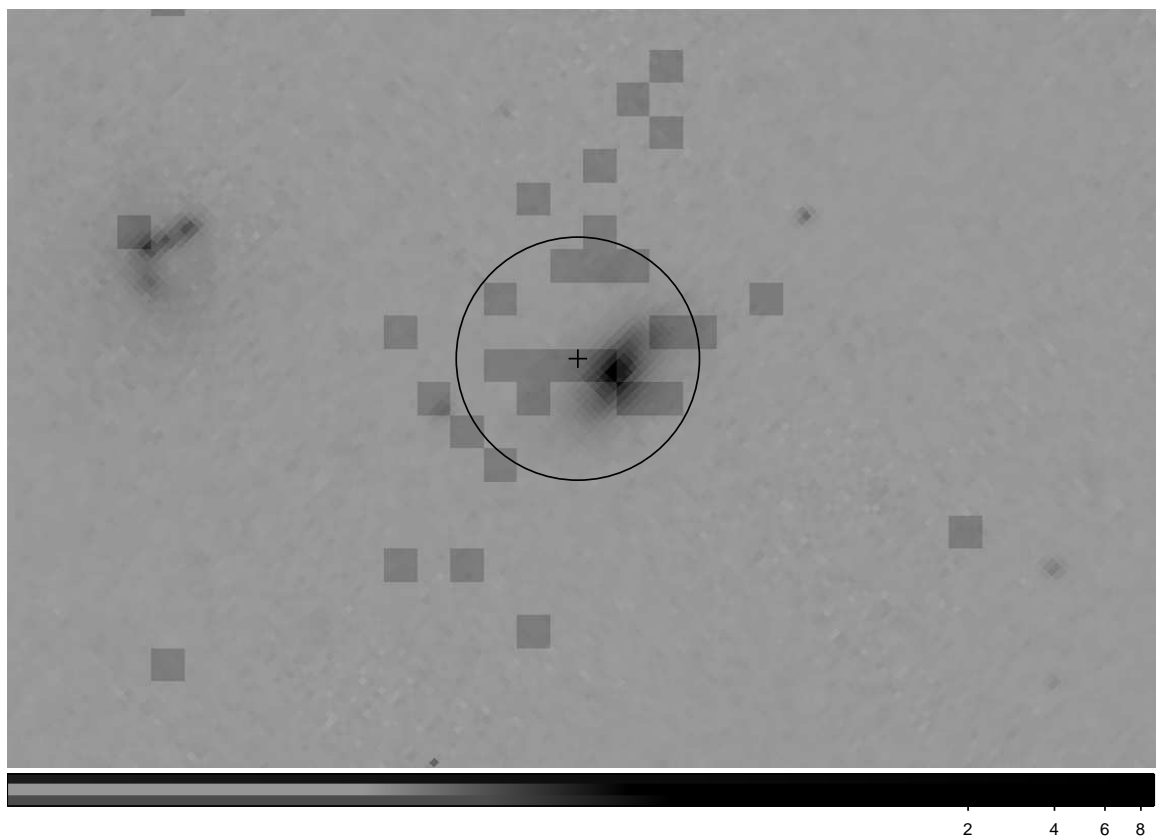}
\caption{Pre-flare HST WFPC2 image of SDSS J131122.15-012345.6 taken
  using the F606W filter.  The larger overlaid pixels are singly binned X-ray events
  from the flare at is peak at epoch 2004.16.  The circle indicates
  centroid error at $r\sim$1\farcs8 and the cross is the middle of
  the centroid.}
\label{hstcxc}
\end{figure}

We retrieved archival images from the Hubble Space Telescope taken
using the Wide Field Planetary Camera.  \object{SDSS J131122.15-012345.6} was
observed as part of \object{Abell 1689} by WFPC2 for multiple $\geq 900$s
orbits on March 2, 1996 and July 4, 1996 using the F606W and F814W
filters, although a single orbit from each filter was sufficient for
our purposes.  To improve relative astrometry, we shifted the HST
image to align with \cha\ coordinates.  We determined the centroid
coordinates of other bright X-ray sources within the WFPC2 FOV and
visually identified the centers of corresponding HST objects that also
had SDSS catalog entries.  SDSS objects corresponded reliably to
\cha\ objects and were much more numerous than \cha\ objects, so we
used additional SDSS objects to confirm HST correspondence.  We find
the centroid of the \cha\ source to be within a \cha\ pixel
($0.\arcse5$) of the host object's center, but the primary source of
positional error is the statistical uncertainty in the centroid of the
X-ray source, $\sim1\farcs8$.  The correspondence of the galaxy to
the \cha\ X-ray source can be seen in Figure \ref{hstcxc}.  Later we will assume the X-ray source is indeed associated with SDSS J131122.15-012345.6 unless explicitly stated otherwise.

We confirmed the host object as a galaxy whose morphology appears plausibly
elliptical but exhibits some spiral symmetry, possibly containing
spiral arms that are either broad or too smooth to definitively
demarcate at the galaxy distance.  Using IDL user-contributed programs
\citep{Landsman93,Barth01} for aperture photometry and fitting the
flux to a \cite{Sersic} $R^{1/n}$ radial profile, we found a
S\'{e}rsic index $n\simeq1.2$, which suggests dominance by an
exponential disk model (see, e.g. \cite{Blanton03,GD05}.  The flux in
excess of an $n=1$ profile lies within $\sim0.\arcse15$ of the center
(a 3x3 grid of detector pixels) and suggests a maximum bulge-to-total
luminosity ratio B/T$=0.18$.  The radial profile is a poor fit to a
deVaucoleurs profile where $n=4$, even excluding the core excess,
suggesting the galaxy is not elliptical.  We further confirm a spiral
disk-and-bulge model by examining the ratio of F606W to F814W as a
function of radius.  The ratio is approximately flat as a function of
radius except within the region of core excess, where it demonstrates
a sudden red spike as might be produced by a spiral galaxy with a red
core and as opposed to a more gradual reddening in an elliptical. We conclude, therefore, 
the galaxy is most likely a spiral. 

\subsubsection{HET}

We observed \object{SDSS J131122.15--012345.6} using the {\it Hobby-Eberly
Telescope} (HET) for a total of 5900 seconds split into 4
sub-exposures.  The first two were taken on March 24, 2009 and the
other two on March 27, 2009, well after we would expect to see any
emission lines due to the tidal flare itself.  We used the LRS
(low-resolution spectrograph) with the G1 grism (300 l/mm) and the
2\arcsec\ slit.  This gave us a resolution of 14.8\AA\ or 800 km/s at
6000 \AA.  The spectrum covered the range 4080 - 10800 \AA, with the
peak sensitivity between 5500 and 6000 \AA.  The signal-to-noise ratio
(hereafter S/N) in this spectrum is approximately 4 at 5000~\AA, 7 at
5300~\AA, approximately flat at a value of 10 between 5500 and
7200~\AA, and drops to values of 3--6 between 7500 and 8500~\AA. At the
longest wavelengths, the noise is dominated by imperfect subtraction
of the strong OH emission lines in the spectrum of the sky background.

\begin{figure}
\includegraphics[angle=270,scale=0.70]{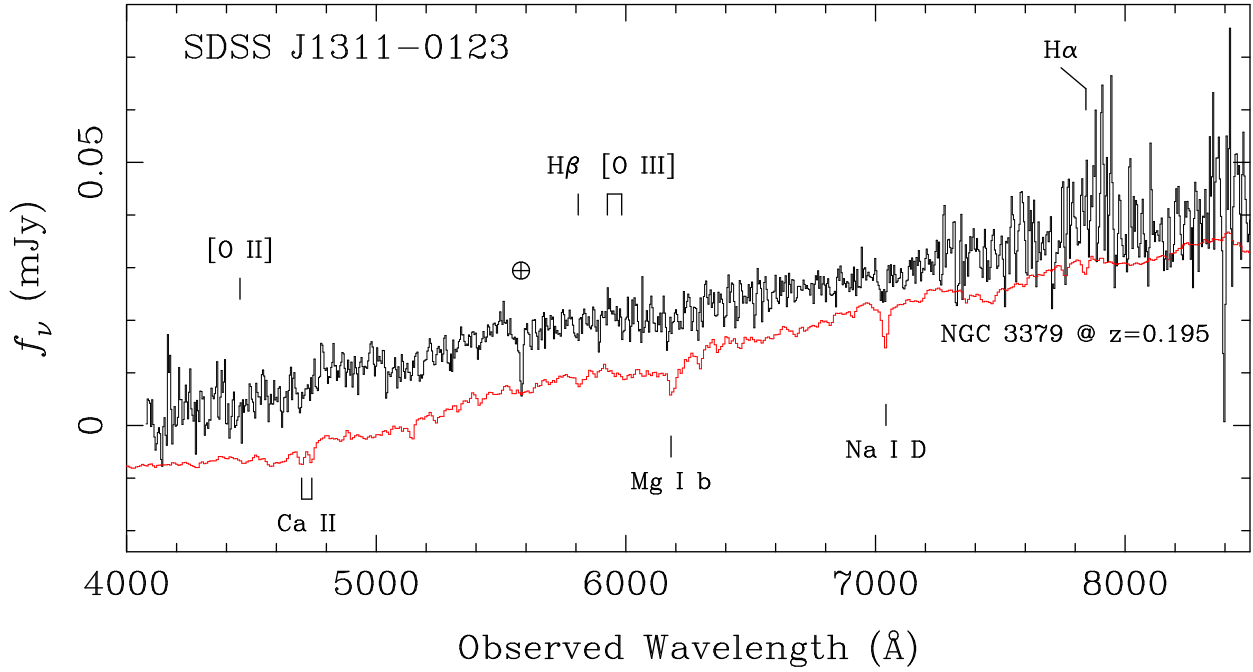}
\caption{Observed {\it HET} spectrum of the flaring galaxy with a spectrum of
the nearby elliptical galaxy NGC 3379 shifted to $z=0.195$ overlaid
for comparison.  There is a residual from the imperfect subtraction of
the sharp [\ion{O}{1}] telluric line at 5577\AA, marked with a crossed
circle.  In addition, there are also large residuals from poor sky
subtraction at wavelengths longer than 7500~\AA. Amongst the galaxy's
absorption features, at 7000\AA\ we see the \ion{Na}{1}\,D
interstellar line, at 4650\AA\ are \ion{Ca}{2} H and K, and at
6200\AA\ is \ion{Mg}{1}b.  The spectrum near H$\alpha$ is dominated by noise from sky subtraction residuals.  Note the lack of strong emission lines that
would indicate a typical AGN.
}
\label{hetspec}
\end{figure}

{\it Absolute} spectrophotometry with the HET can suffer from large
errors because a variable fraction of the light incident on the
primary mirror is focused onto the secondary as it tracks the object
across the filed of view. Therefore, we have adjusted the absolute
flux scale by in order to match the SDSS photometry. After diving the
original flux scale by 2.75 we achieve a very good agreement with the
$r$, $i$, and $z$ magnitudes. The observed spectrum is shown in
Figure~\ref{hetspec}. By comparison with the spectrum of the nearby
elliptical galaxy NGC~3379, We identify features that we interpret to
be the \ion{Na}{1}~D interstellar double near 7000\AA\ and the
\ion{Ca}{2} H and K lines near 4650\AA\ at target-redshifted
wavelengths, as well as \ion{Mg}{1}~b absorption near 6170\AA. Thus,
we find $z=0.195$, in good agreement with photometric results from
$LePhare$. These features are marked and labeled in
Figure~\ref{hetspec}.  The signal-to-noise ratio is insufficient to
eliminate ambiguity from the redshift estimate, in part due to the
lack of strong emission lines.  This redshift is supported by template fitting with the {\it EZ} tool that was used in the VIMOS VLT Deep Survey \citep[VVDS,][]{VVDS05} and includes spectral line identification capabilities.  The best fits are for elliptical templates which also confirm absorption in \ion{Ca}{2} and \ion{Mg}{1}~b, as well as the G band, H$\gamma$ and H$\beta$ near 5139\AA, 5186\AA\ and 5809\AA.  A starburst model also reproduces this redshift but with a continuum significantly bluer than observed.  None of the {\it EZ} AGN models converged at the redshift of the object (V. Le Brun 2009, private communication).

In order to evaluate the possibility that emission lines form an AGN
are hiding in the optical spectrum, we have determined upper limits on
the flux of some of the relevant lines. In particular, we have focused
on the observed wavelength range 5800--6000~\AA, where we expect to
find the H$\beta$ and [\ion{O}{3}]~$\lambda\lambda$4959,5007 emission
lines (marked an labeled in Figure~\ref{hetspec}). This also happens
to be the region of the spectrum where the S/N is highest. Although
the H$\alpha$ line is stronger than H$\beta$, it falls at an observed
wavelength of 7843~\AA, where the noise is very high. The positive
spikes discernible in this part of the spectrum are a result of poor
subtraction of emission lines from the night sky. We expect the
[\ion{O}{3}] lines to be unresolved, therefore, we assume a FWHM of
800~km~s$^{-1}$ according to the spectral resolution attained in the
spectrum. Under this assumption we obtain an upper limit to the
observed flux of the [\ion{O}{3}]~$\lambda$5007 line of $1\times
10^{-16}~{\rm erg~s^{-1}~cm^{-2}}$. If a line is stronger than this
limit, its peak stands out by more than $3\sigma$ above the local
noise. In the case of the H$\beta$ line, we considered two different
values of the FWHM, 900~km~s$^{-1}$ (marginally resolved) and
1600~km~s$^{-1}$. We chose these particular values because they are
typical of the H$\beta$ lines of Narrow-Line Seyfert 1 galaxies
(hereafter NLS1s), which are known for their large-amplitude X-ray
variability  \citep[e.g.][]{Boller93,Boller97,Grupe95a,Grupe08}. Thus the limits we obtain under these assumptions will be
relevant to our discussion of whether the observed flare could be an
example of dramatic variability in a NLS1 galaxy. For a FWHM of
900~km~s$^{-1}$ the upper limit to the H$\beta$ is $1\times
10^{-16}~{\rm erg~s^{-1}~cm^{-2}}$ (just as for the
[\ion{O}{3}]~$\lambda$5007 line), while for a FWHM of 1600~km~s$^{-1}$
the upper limit to the H$\beta$ is $2\times 10^{-16}~{\rm
erg~s^{-1}~cm^{-2}}$. For line broader than this, the upper limit to
the integrated flux increases proportionally to the FWHM.

\subsubsection{GALEX}

Ultraviolet observations of SDSS J131122.15-012345.6 using the {\it Galaxy Evolution Explorer}\footnote{http://www.galex.caltech.edu} ({\it GALEX}) exist for the dates of 22 April 2004 and 24 April 2007.  These $\sim100$s survey observations have limiting magnitudes of $\sim20-22$ for bands about $\sim1500$\AA\ and $\sim2300$\AA.  No emission is obviously visible by inspection at the location of the galaxy in either band, whether from the galaxy as a  whole or a point within the galaxy.  But given the limiting magnitude of these short observations, they should be insufficient to detect even an unabsorbed $0.12$ keV blackbody emitting $\lesssim4.7\times10^{-14}$ \ecmss, as derived from the \cha\ spectrum of 28 February 2004.  Furthermore, the galaxy lies at $\sim3\arcmin$ of either field's edge and is thus flagged for instrumental issues due to rim proximity.

\section{Discussion}

\subsection{Observational and Theoretical Background}

In order to place our new results in context, we first provide brief
reviews of both previous observations and estimates of tidal flare
rates.

Although we expect tidal disruption flares to be among the most
luminous observable astrophysical events, with the total kinetic
energy of ejected debris exceeding that of supernovae at $10^{51}$ erg
or more, these flares have thus far been challenging to observe.
While they should occur within AGNs and may contribute significantly
to the faint end of the AGN X-ray luminosity function \citep{Milos06}, they
will be easiest to identify in a quiescent galaxy, where they can be
distinguished from typical variable disk accretion in an AGN.

Existing theoretical studies predict disruption rates of 1 event per
$10^4 - 10^5$ years per galaxy, \citep[e.g][]{MT99}, a rate that has been
supported by observational studies using ROSAT \citep{Donley02} . The
most optimistic predictions increase that rate by an order of
magnitude \citep{Wang04}.

Several candidate events have been observed \citep{Komossa05,
  Komossa04, Halpern04}, but some of the most convincing evidence of
tidal disruptions comes from \galex\ detections of UV flares with
optical and X-ray component as observed by
\cite{Gezari06,Gezari08,Gezari09}.  Ongoing studies using the XMM Slew
Survey \citep{Esquej07,Esquej08} also report candidate events, but an
extensive study of the Chandra Deep Field \citep{Luo08} made no
detections, consistent with maximum rates comparable to $10^{-4}$ per
galaxy per year for $L \gtrsim 10^{43}$ \es.  Since only about a dozen
such candidate events have been identified to date (see the above
references), the statistical conclusions that can be reached thus far
are highly tentative.

The rate by which tidal flares occur should also act as an indicator
for the distribution of black holes in the galaxy population.  The
effect may be particularly pronounced according to the theoretical
calculations of \cite{Wang04} if a significant fraction of nucleated
dwarf spheroidal galaxies harbor MBHs at their centers.  Given that dwarf
ellipticals are a very numerous component of the galaxy distribution
\citep[see, for example][]{Jenkins07}, if lower mass MBHs flare more often than more massive MBHs, they may dominate the flare rate if they contribute at all.  But more recent work by
\cite{Merritt09} suggests lower mass MBHs may produce such flares more
rarely even if dwarf galaxies do possess MBHs.  As noted
in the introduction, determining the population of MBHs in dwarf
spheroids through such indicators as tidal disruption events will also affect
predicted rates of MBH-MBH mergers and extreme mass ratio inspirals
(EMRIs).  Although the gravitational wave detector {\it LISA} will be sensitive only to disruptions of stars with compact cores \citep{ASetal07}, as indicators of the dwarf galaxy MBH population 
these predicted rates will have additional implications for {\it LISA}, which will be sensitive to deeply penetrating encounters in the mass range of $\Mbh \lesssim 10^{7}\;\Msun$
\citep{Jennrich04,SHMV04,SHMV05,Sigurdsson03}.  

\subsection{Derived Galaxy Properties}

\subsubsection{Cluster Membership} 

With a spectroscopic estimate of $z=0.195$ for SDSS
J131122.15-012345.6, we take the galaxy to be a member of \object{Abell 1689}.
A variety of redshift estimates have been published for \object{Abell 1689}
ranging from 0.183 \citep{Struble99} to 0.203 \citep{Estrada07}.  The
high velocity dispersion of this massive cluster suggests that
membership of the SDSS galaxy is plausible even for a low estimate of
the cluster redshift.  \cite{Halkola06}, for example, estimate a
velocity dispersion $\sigma\sim1500\ $km s$^{-1}$, placing the galaxy
in the upper $\sim1\% $ of cluster members for the low $z$ estimate.
\cite{Duc02} define an inclusive cluster range of $z=0.17-0.22$ using
both photometric and spectroscopic redshifts.  And \cite{Martini07}
estimate $\sigma=2402\ $km s$^{-1}$ for $z=0.1867$.  For the purpose
of calculating luminosities then we select a characteristic redshift
$z=0.19$ for which the luminosity error will be a factor of less than
1.09 whether or not the SDSS galaxy's higher measured redshift
signifies Hubble expansion rather than motion relative to the cluster
center.  In our adopted cosmology, $z=0.19$ corresponds to a luminosity distance of 926 Mpc.

\subsubsection{Black Hole Mass}

Since we do not have a measurement of the stellar velocity dispersion
of the host galaxy, we have used photometry to estimate the mass of
its central black hole (assuming it has one).  We estimated \Mbh\ from M$_V$ as per
\cite{Lauer07}, as well as from $M_B$ as per \cite{FF05}.  We
determine M$_V$ and $M_B$ from SDSS $ugriz$ magnitudes and the
conversion method prescribed by \cite{Lupton05}.  We specifically use
the calculations that rely upon the $g$ and $r$ bands only given the
likely unreliability of $u$.  We have not applied K-corrections as the
adjustments they introduce at $z=0.19$ will be minor compared to the
errors resulting from scatter in the photometric $\Mbh-L$
relation. Using the ratio of bulge luminosity to total galaxy
luminosity (B/T) based on the SDSS data, we find for the bulge
M$_V=-17.10\pm0.10$ and M$_B=-15.98\pm0.10$.  

We then refer to \cite{Lauer07} and \cite{FF05} to estimate \Mbh .  Given M$_V=-17.10$, we use the \cite{Lauer07} relation that includes their entire dataset, and is hence appropriate for M$_V>{-19}$.  This relation implies log$(\Mbh/\Msun)=6.23\pm0.29$, and the \cite{FF05} relation produces log$(\Mbh/\Msun)=6.68\pm0.24$.  Taking the
mean from these relations we find log$(\Mbh/\Msun)=6.46\pm0.38$, an estimate that is comfortably compatible with the possibility of tidal disruption events, even if a much larger error were inferred due to the inherent uncertainty in estimating (B/T).  The indicated error arises primarily from scatter in the relations we reference, but our ability to measure
(B/T) is limited by WFPC2 pixel scale.  For example, the host galaxy could have a more compact bulge and hence a lower inferred black hole mass.  We therefore determine the largest uncertainty in the lower bound of \Mbh\ by assuming that the bulge is described by a single pixel.  In this case, the additional uncertainty due to (B/T) is 
$\Delta$log$(\Mbh/\Msun)\propto
log ($B/T$)\sim0.4$.

\subsection{Was the Event a Tidal Flare?}

In the following subsections we argue that alternative explanations of
the observations are disfavored by the data. We then demonstrate that
the data are consistent with the event being a tidal flare.

\subsubsection{A Galactic Foreground Object?} 

At a galactic latitude of $61\degr$, the position of \object{Abell 1689}
already produces a relatively low probability of having a
line-of-sight foreground object e.g. a distant M-star.  We now
consider the following: (1) A source within the galactic plane, at a
distance of $\sim1$ kpc, would have $L_X$(0.5--2.5 keV) $\lesssim10^{30}$ \es, whereas
known quiescent low-mass X-ray binaries (qLMXBs) typically have
$L_X$(0.5--2.5 keV) $>10^{31}$ \es\ and hard spectra, $\Gamma\sim1-2$ \citep{Heinke03}.  (2) M-dwarfs are
considerably more common and occasionally produce X-ray flares, but
even a faint galactic M-dwarf should be visible against our proposed
host galaxy for the event in HST image (take, for example, the M-dwarf
selection criteria of $r<19.0$ by SEGUE, \citep{SEGUE09}).

\subsubsection{A Highly Variable AGN?}

We consider alternate extragalactic explanations for the X-ray flare
due to known types of X-ray emitting objects.  For example, AGNs may
show X-ray variability of a factor of a few but typically not bright
outbursts of $\gtrsim30$.  Characterization of the X-ray continuum and
the observed power law of $\Gamma \gtrsim5$ is also atypical of
AGNs, which are better described by $\Gamma \lesssim2.5$
\citep[e.g.][]{Beckmann06}, save for NLS1s.  Although photometric fitting is
consistent with a QSO at $z\sim3.4$, luminous AGNs typically have
prominent emission lines at optical rest wavelengths whereas the host
shows no strong emission lines in the HET spectrum with limits as
reported in \S2.3.4.  A template fit requiring a photometric redshift
$z\sim3.4$ is also contraindicated by the HET redshift, $z=0.195$.
Given the optical spectral information, we judge therefore that an AGN
is unlikely to have caused of the observed flare, with the following
two exceptions: BL Lac objects and Narrow-Line Seyfert 1 galaxies
(NLS1s, known for their steep soft X-ray spectra and rapid X-ray
variability). We consider these two possibilities in turn.

{\it A BL Lac Object. --} There exists no corresponding radio
source in the FIRST catalog \citep{FIRST97} as would be expected for a
BL Lac.  \citep[For a review of BL Lac properties, see][]{Urry95}.
The FIRST catalog reaches a sensitivity 0.75 mJy at 1.4 GHz and
includes the \cha\ fields.  Given that most BL Lac objects have
elliptical galaxies as hosts, the favored spiral classification of the
host galaxy also implies the flare does not arise from a BL Lac
object.

{\it A Narrow-Line Seyfert 1 Galaxy or Similar Object? --} There are a
number of examples of NLS1 galaxies displaying dramatic X-ray
variability (with amplitudes of more than a factor of 10). These
include WPVS~007, which showed a drop in its soft X-ray flux by a
factor of $\sim 400$ in 3 years \citep{Grupe95a}, and Mrk 335,
which showed a drop in its soft X-ray flux by a factor of $\sim 30$ in
approximately 1.4 years \citep{Grupe08}. X-ray variability by a
factor of up to 10 on time scales as short as days is a trademark of
NLS1s \citep[see, for example, the light curves of NGC~4051 in ][]{McHardy04}. An extreme example of this type of behavior is the
variability of IRAS~13224--3809 \citep{Boller93,Boller97}. There are
also examples of dramatic X-ray flares in AGNs that are not NLS1s,
including IC~3599 (also known as RE~J1237+264 or Zwicky~159.034) and
NGC~5905. Both of these are Seyfert 2 galaxies with X-ray excursions
by factors of 120 and 150 respectively \citep[see ][]{Grupe95b,BPF95,BKD96,KB99,Gezari03}.  

The characteristic broadband spectral properties of SDSS~J131122.15-012345 are generally not unheard of among NLS1s, but they are atypical in several respects.  Like NLS1s in general, its X-ray emission is very soft.  The vast majority of NLS1 X-ray spectra, however, can be fit with power laws such that $\Gamma\sim2-4$ \citep{Grupe04}.  Both high and low states of SDSS~J131122.15-012345.6 are fit to power laws with $\Gamma\sim5$, which is comparable to the softest examples from \cite{BPF95} and moderately steeper than these examples of dramatic Seyfert flares, with the dramatic exception of WPVS~007 ($\Gamma\sim10$) which has been attributed to the presence of a warm absorber \citep{GLK08}.  The correlation between spectral slope and $L_X$ observed by \cite{Grupe04} is more problematic for classifying SDSS~J131122.15-012345 as an NLS1.  The softest object in that sample corresponds to $L_X\sim10^{44}$ \es, $\sim4\times$ brighter than the observed high state power-law $L_X$(0.2-2.0 keV) of SDSS~J131122.15-012345.  This high state $L_X$(0.2-2.0 keV) conversely corresponds to $\Gamma\sim2.5$ in the \cite{Grupe04} sample, compared to the observed $\Gamma\sim5$.  The low state $\Gamma$ is similarly steep, consistent with no time evolution, and more difficult to reconcile with this $\Gamma-L_X$ relationship.  Examining the properties of the optical continuum, we find that the SDSS $g$-band absolute magnitude $M_g=-18.4$ is low for the sample examined by \cite{Zhou06}, who find that soft optically selected NLS1s almost exclusively occupy $-19\lesssim M_V\lesssim -25$.  This $M_g$ corresponds to a HET spectrum that declines steadily with decreasing $\lambda$ for $\lesssim5500$\AA\ (SDSS color (u-g)$\sim3.8$), which would make a similar NLS1 rare and "UV-deficient" compared to a typical flat-spectrum NLS1 in \citep{Zhou06}.

To investigate whether the observed flare in SDSS~J131122.15-012345.6 is consistent with the variability
of an AGN, we ask whether optical emission lines characteristic of an
AGN could be hidden in its optical spectrum below our detection
limit. To answer this question, we have studied the relation between
the 0.2--2.0~keV X-ray luminosity and the (broad) H$\beta$ and
(narrow) [\ion{O}{3}]~$\lambda$5007 line luminosities among the 51
NLS1 galaxies in the sample of \cite{Grupeetal04}. In this sample,
the ratio $F_X({\rm 0.2-2\,keV})/F({\rm H\beta})$ has a median value
of 127, with 90\% of objects having $F_X({\rm 0.2-2\,keV})/F({\rm
H\beta}) > 50$. Similarly, the ratio $F_X({\rm 0.2-2\,keV})/F({\rm
[O\,III]})$ has a median value of 486, with 90\% of the objects having
$F_X({\rm 0.2-2\,keV})/F({\rm [O\,III]}) > 100$ Using the best-fitting
power-law model for the spectrum of SDSS~J131122.15-012345.6, we find
that in the quiescence $F_X({\rm 0.2-2\,keV}) < 5.1\times
10^{-15}~{\rm erg\; cm^{-2}\; s^{-1}}$, implying that $F({\rm H\beta})
< 1\times 10^{-16}\; {\rm erg\; cm^{-2}\; s^{-1}}$ and $F({\rm
[O\,III]}) < 5 \times 10^{-17} \; {\rm erg \; cm^{-2} \;s^{-1}}$ (at
90\% confidence), which fall just below our detection limit.  However,
we doubt that this object can be a NLS1 galaxy because the upper limit
on its quiescent X-ray luminosity suggests that any AGN should be
accreting at a rate well below the Eddington limit, contrary to
NLS1s. In particular, using the \cite{Elvis94} spectral energy
distribution as a template, the upper limit on the quiescent X-ray
luminosity, and the black hole mass derived above, we infer an upper
limit on the Eddington ratio in quiescence of $L_{\rm bol}/L_{\rm Edd}
<  0.01$. In contrast NLS1 galaxies have $L_{\rm bol}/L_{\rm
Edd} > 0.1$ \citep[e.g.][]{MG05}. Considering now the cases of
the Seyfert 2 galaxies, IC~3599 and NGC~5905, we notice that {\it in
their high X-ray flux states} they had $F_X({\rm 0.2-2\,keV})/F({\rm
[O\,III]})=240$ and 10 respectively. If these values are
representative and if SDSS~J131122.15-012345.6 is a similar object, we
would expect it to have $F({\rm [O\,III]})\approx(1-30)\times
10^{-15}\; {\rm erg \; cm^{-2} \;s^{-1}}$ (since it has $F_X({\rm
0.2-2\,keV})=2.7\times 10^{-13}\; {\rm erg \; cm^{-2} \;s^{-1}}$ for a power law in
its high state). These line fluxes are an order of magnitude above our
detection limit, yet we have not detected any lines, which leads us to
conclude that SDSS~J131122.15-012345.6 is unlike IC~3599 and NGC~5905.
Rather it resembles RX~J1242.6-1119 \citep{KG99} and
RX~J1624.9+7554 \citep{GTL99}, which are apparently
normal galaxies that exhibited large, soft X-ray flares with
amplitudes of 20 and 235, respectively.  Of course, without additional
observational constraints near the flare peak we cannot entirely
exclude the possibility that, despite having excluded known modes of
AGN emission and variability, we have found an extremely rare instance
of supersoft variability from a quiescent or obscured AGN and not a
stellar tidal disruption.  Explanations might include a novel mode of
fluctuation in an otherwise low-luminosity (perhaps radiatively
inefficient) accretion flow or a temporary gap in obscuring material.

\subsubsection{Other Extragalactic Line-of-Sight Objects?}
	
As described in \S2.1, supernovae have been observed rarely to emit
prompt X-ray flashes that reach luminosities comparable to those
observed in the candidate flare, such as from a progenitor shock
breakout \citep{Campana06,Soderberg08,GezariSN08}.  In all cases, the
soft thermal component that dominates this prompt flash evolves
rapidly over timescales of less than the 20~ks peak observation of our
candidate flare.  This evolution appears to occur in conjunction with
a softening of $kT_{bb}$ as the breakout expands into larger distances
from the progenitor surface.  The A1689 flare demonstrates no such
short-term variability during the peak epoch, excluding a progenitor
breakout as an explanation.  The flare is also orders of magnitude
more luminous than even more luminous examples of later, long-term
evolution of supernova ejecta \citep{Immler03,SP06,Immler08},
suggesting that a different explanation is also necessary here.

X-ray afterglows from gamma ray bursts (GRBs) have been observed on
time scales of years \citep{Grupe10} but the decay is sufficiently
rapid (declining a factor of $10^5$ within $10^{5}$ s of peak X-ray
luminosity, then another factor of $10^5$ by $10^8$ s) and the X-ray
spectrum sufficiently hard ($\Gamma\sim1$) that we reject them as
explanations for the event we detected.

\subsubsection{Tidal Flare Explanation}

We now compare our result to previously reported events.  Early \cha\
observations of SDSS J131122.15-012345.6 indicate no detectable source for observations of at epochs 2000.29 and 2001.98 with
$L_X$(0.3--3 keV) $<1.6\times10^{41}$ \es\ at $z=0.19$.  The \xmm\ observation places a similar limit of
$1.1\times10^{41}$ \es\ at epoch $2001.98$.  Then, we detected a prominent $\sim11\sigma$
super-soft flare with an extinction-corrected $L_{X}$(0.3--3 keV) $=4.8\times10^{42}$ \es\ at
$t=2004.16$, as well as evidence ($\sim 3\sigma$) of emission two years later at $t=2006.18$ with
$L_{X}$(0.3--3 keV) $=5.2\times10^{41}$ \es.  At $t=2007.18$, $L_{X}$(0.3--3 keV)
$<3.6\times10^{42}$ \es.  
Unless otherwise indicated, further values of $L_{X}$ for this flare
will use a value as determined from a 0.12 keV blackbody in the 0.3--3 keV
band, corrected for extinction through the plane of our galaxy as per \S 2.2.3.  

The peak observed
X-ray luminosity and temperature are as predicted by theory and
typical for X-ray observations of tidal flare candidates to date (see
\cite{Gezari09} for a review and refer to their Figure 10), and the
rise in luminosity by a factor of $\sim30$ over pre-flare background
is typical but not remarkable.  For example, \cite{Esquej08} consider
candidates with $L_X$(0.2--2 keV) variable by $\times 20$ and \cite{Luo08} require
a count rate variable by $>20$, but we see a factor of $\gtrsim650$
in $L_X$(0.3--2.4 keV) for \cite{Cappelluti09}.  

The observed $L_X$ never approaches theoretical peak luminosities of $>10^{45}$ for a tidal flare \citep{Ulmer99}. However, the flare may have been observed
significantly past its peak or could be the result of a ``weak''
encounter that stripped the outer layers of the stellar atmosphere.  Our ability to estimate the flare's bolometric luminosity ($L_{bol}$) is also limited by the spectral response of ACIS-I below $\sim0.5$ keV.  If the spectrum is better described as a steep absorbed power law or multi-temperature blackbody, $L_{bol}$ could be an order of magnitude or more greater than $L_X$.

{\it We thus find a tidal disruption event located at the center of SDSS J131122.15-012345.6 to be the most likely
explanation for the X-ray flare and the following discussion proceeds
based on this assumption.}

\subsection{Comparison of Observations to Tidal Flare Models}

With only a few data points we cannot determine the shape of the
flare light curve observationally.  But by examining the flare's X-ray light curve and spectral properties, we show here that the flare is consistent with the expected behavior of a tidal disruption event.

\subsubsection{Light Curve Decay}

\paragraph{Comparison with a Tidal Flare Accretion Model:}

At the early times, the mass infall rate $\dot{M}$ for fallback of the stellar debris from a tidal disruption may exceed the Eddington limit $\dot{M}_{Edd}$, even as the luminosity flare remains near the Eddington luminosity $L_{Edd}$ \citep[e.g.][]{Ulmer99}.  The central UV/X-ray source may also be obscured at the earliest times by optically luminous super-Eddington outflows on a timescale of days \citep{SQ09}.  

At later times, when $\dot{M}\lesssim\dot{M}_{Edd}$, the luminosity is expected to track the
rate of return of stellar debris and the accretion rate onto the black
hole with a decline according to $[(t-t_D)/(t_0-t_D)]^{-n}$ where
$n=5/3$ by theory \citep{Phinney89,EK89,ALP00}, $t_0$ is the peak $\dot{M}$ and $t_D$ is the time of disruption.  

The observations constrain
$t_0$ between the \xmm\ nondetection at 2001.98 and the \cha\
high state at 2004.16.  Even though there are several archival
observations of Abell 1689, the flare was clearly detected at only only
two epochs and was barely detectable by 2006.18.  Given the short
duration and the resulting low level of statistical significance of
the 2007.18 observation, we omitted it as a constraint for the purposes
of fitting the decay of the light curve and simply solve for $t_D$
analytically with two points.  For a time $t$ in years where $t>t_0>t_D$
and $t_D\sim2003.40$, we obtain

\begin{eqnarray}
L_X(0.3-3~\textrm{keV})=(3.0\pm2.9)\times10^{42}\ \textrm{erg s}^{-1}\nonumber\\*
\times\left[\frac{t-(2003.40\pm0.49)}{1 \textrm{ yr}}\right]^{-5/3}.
\label{decayeq}
\end{eqnarray}

\begin{figure}
\plotone{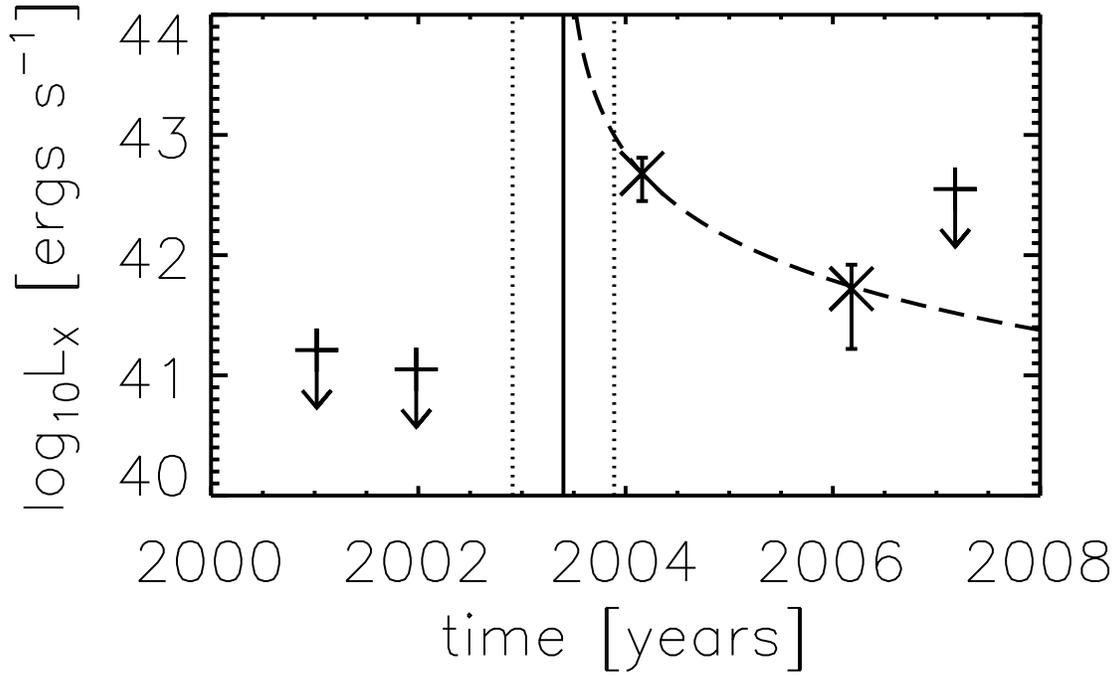}
\caption{Light curve for the flare with error bars.  Arrows indicate
  upper bounds.  $L_X$(0.3--3) corrected for galactic extinction are indicated by $\times$.  The dashed
  line is a $t^{-5/3}$ light curve which assumes $t_D$ at
  the expected value for a solar-type star, indicated by the solid vertical line.  Dotted
  lines indicate the error range for $t_D$.  The rise in luminosity
  between $t_D$ and $t_0$ is not indicated, nor is the peak luminosity
  at $t_0$ shown.
}
\label{lc}
\end{figure}

This relation is shown in Figure \ref{lc} with data points, error bars
and upper limits included.  The expected rise from $t_D$ to $t_0$ is
not indicated but is discussed below.  Note the large uncertainty in
the normalization is strongly correlated with the uncertainty in
$t_D$.

The variability and spectral shape (see \S2.2.2, \S2.2.3, and Figs. \ref{cr_vs_hr} and \ref{xspec}) of the X-ray data are therefore consistent with the interpretation that we have identified a tidal disruption event, which we will examine in more detail as follows.

\paragraph{Dynamical Interpretation:}

By examining plausible values for the dynamical timescale of the disruption, we place constraints on both the time of disruption and the properties of the disrupted star.
Theoretically, the difference between $t_D$ and $t_0$ should be governed by the specific energy of the stellar
material falling back to the black hole, such as is calculated by
\cite{LNM02} as follows.  Let \Rstar\ and \Mstar\ be the radius and
mass of the disrupted star, $R_p$ is its periastron and $R_t$ is the
tidal radius of the MBH with mass \Mbh. Then,

\begin{eqnarray}\label{li02tscale}
(t_0-t_D)=0.11\textrm{ yr }k^{-3/2}\left(\frac{\Mbh}{10^6\;\Msun}\right)^{1/2}\nonumber\\*
\times\left(\frac{R_p}{R_t}\right)^3\left(\frac{\Rstar}{\Rsun}\right)^{3/2}\left(\frac{\Mstar}{\Msun}\right)^{-1}(1+z).
\end{eqnarray}
We have increased the timescale here according to redshift of the host galaxy, $z$, as per \cite{Gezari09}.
The value of $k$ depends on the extent of the star's spin-up by the tidal encounter, with a
minimum of $k=1$ for a non-rotating star and maximum $k=3$ for a star
that is spun up near the point of break-up.  Simulations and linear analysis favor values approaching maximum value, although the full range is possible in principle \citep{Alexander01,LNM02}.

If log$(\Mbh/\Msun)=6.46\pm0.38$ for the host galaxy, as derived from the $\Mbh-L$ relationship, then for a
main sequence star and $k=3$, 
\begin{equation}
0.02<(t_0-t_D)\times(R_p/R_t)^{-3}<0.06\ \rm{yr}.
\end{equation}
This short dynamical timescale implies $t_0 \ll 2004.16$ for a main sequence star, suggesting the flare is consistent with a tidal disruption event which is already well decayed from its peak when first observed.


The analysis above assumes $\Mstar\sim\Msun$ and $\Rstar\sim\Rsun$, but should also be appropriate for a wide range of plausible stellar masses and radii.  Note that $(t_0-t_D) \varpropto (\Mstar/\Msun)^{-1}
(\Rstar/\Rsun)^{3/2}$, and while it is common to assume a stellar
mass-radius relationship \Rstar=\Rsun(\Mstar/\Msun)$^{\alpha_R}$ where
$\alpha_R=1$ for main-sequence stars of $\Mstar<1\, \Msun$
\citep[e.g.][]{LNM02}, $\alpha_R\sim0.75$ is empirically appropriate
for main sequence stars of $\Mstar>1$ \citep{HKT}.  Above $\Mstar=1\,
\Msun$, $(t_0-t_D)$ depends only weakly on increasing stellar mass.
This includes the characteristic properties of luminous S-stars such
as dominate observations of our own galactic center
\citep{Alexander05} and are therefore of great interest as plausible tidal disruption progenitors.

Suppose an alternative interpretation that does not presuppose unobserved early peak emission, such that $t_0\sim2004.16$.  Such a scenario would be possible if the disrupted star had a large radius, such as a giant, facilitating very long decay timescales \citep[e.g.][]{Lodato09}.  The disruption of a giant is plausible given that observations of our own galaxy have demonstrated the presence of numerous early-type giants, as well as a depleted population of late-type giants, within $\sim0.1$ pc of Sag A$^*$ \citep[e.g.][]{Do09}.

In summary, we find that the dynamical time scale of the light curve implies a flare significantly decayed from its peak luminosity or that the disrupted star was a giant.  The data do not allow us to distinguish between these two possibilities.

\subsubsection{Accreted Mass}

To further examine the properties of the disrupted progenitor, we estimate the total X-ray energy released by integrating the
luminosity over the duration of the flare.  The total mass accreted is
thus

\begin{equation}
\Delta M=\frac{\Delta E}{\epsilon c^2}=\frac{f_{bol}}{\epsilon c^2}
\; \int_t^\infty L_X(t)dt,
\end{equation}

\noindent where  $\epsilon\sim0.1$ and $f_{bol}$
is the bolometric correction (here $\sim1.4)$.  We derived the bolometric correction by taking
the ratio of luminosities over 0.3--3 keV and over all significantly contributing energies as
derived through numerical integration of a 0.12 keV blackbody, for which the peak photon
energy $\sim0.6$ keV.  The bolometric correction could in principle be an order of magnitude larger, however, if instead the X-ray emission arises from a steep absorbed power law or multi-temperature blackbody \citep[e.g.][]{Gezari09,SQ09}.  

For a lower bound on $t$, we could
conservatively use observed luminosity in the high state as the peak
of flare, which is only $\sim1\%$ of the Eddington Luminosity
$L_{Edd}=1.3\times10^{45}(\Mbh/10^7\Msun)\ $erg s$^{-1}$.  This gives
$\Delta M > 1.3 \times10^{-3}\ \Msun$ after the date of peak observed luminosity.  But if $0.02<(t_0-t_D)<0.06$ yr, as predicted from the host galaxy and the \Mbh--$\sigma$ relation, then $\Delta M\sim0.01\ \Msun$ total, such that $\Mstar\simgreat 0.1\ \Msun$ for a mass accretion fraction of 0.1.  A wide variety of conditions could produce modest accretion from a more massive star.  Such scenarios include partial stripping of a star's outer layers or tidal detonation due to a strong counter \citep{CL82,BL08,Esquej08}.  The range of plausibly accreted masses is therefore consistent with the mass range derived in the previous section.

\subsubsection{Geometry of Emission}
 
Finally, we use the X-ray spectral properties to address the plausibility of an emission region described by a disk formed by the accretion of tidal debris. 
 
Following \cite{LNM02}, we can estimate a characteristic radius ($R_X$) of the
region emitting X-rays by requiring the blackbody radiation arise from
a projected disk and assuming a correction factor $f_c=3$ to the
blackbody temperature $T_{bb}$ to account for spectral hardening by
comptonization.  We thus find

\begin{equation}
R_X=\left(\frac{f_{bol}f_c^4L_X}{\pi\sigma T_{bb}^4}\right)^{1/2}.
\end{equation}

\noindent
By this calculation, $R_X=1.3\pm0.4 \times10^{12}$ cm.  For comparison,
the Schwarzschild radius of a black hole
$R_S\sim3\times10^{12}(\Mbh/10^7\Msun)$ cm.  With only $\sim30$ source counts, the flare data do not warrant detailed accretion disk modeling.  
But to first approximation we see that the area of emission is comparable to the estimated size of the host galaxy black hole, $R_S\sim8.6\times10^{11}$ cm.  

If we assume the X-rays trace the innermost edge of an accretion disk, our estimated range of \Mbh\ from the host galaxy also corresponds roughly to a range of innermost stable circular orbits (ISCO) for physically permissible black hole spins $a$, such that $0.62\lesssim R_X/R_S \lesssim 3.6$ \citep[compared to $0.5< R_{ISCO}/R_S < 3$ for $1>a>0$, see for example][]{ST86}.

\subsubsection{Summary of Model-Data Comparisons}

The spectral properties and light curve of the flare are consistent with those predicted for a tidal disruption event of a star by a black hole with \Mbh\ predicted by the $\Mbh-$L relationship.  If described by the characteristic $t^{-5/3}$ decay of a tidal flare, the baseline dynamical timescale of the galaxy's central black hole implies either a disrupted main sequence star that has decayed significantly from its peak luminosity by the time of the observed high state, or a giant star that may instead be near to its peak.  By integrating the light curve of the flare, we find that reasonable assumptions of mass accretion rates and efficiency are compatible with the disruption of a star where $\Mstar \simgreat 0.1\Msun$.  And a blackbody model for the observed spectrum predicts a region of emission consistent with an accretion disk, whose inner edge can be described by the ISCO over a plausible range of \Mbh\ and black hole spin $a$.

\section{Tidal Disruption Rate from Abell 1689}

A comprehensive estimate of the disruption rate per galaxy per year is
complicated by the nonuniformity of the X-ray field coverage,
observation spacing, limiting flux as a function of position the
expected number of galaxies within the X-ray fields of view, and the
expected black hole mass function for different observed host galaxy
types.  A straightforward method of estimating the rate is to
assume that a given number of events $N_t$ is characteristic of the
number of galaxies observed $N_{gal}$ and the timespan $\tau$ over which the observations are sensitive to tidal flare emission.
Let the rate per galaxy per year then simply be
$\gamma=N_t/(N_{gal}\tau)$

%
%

We will first proceed with the conservative assumption that the \object{SDSS J131122.15-012345.6} flare is the only true tidal flare.  This assumption effectively gives us a "best estimate" for the flare rate such that $N_t=1$.
We can estimate the limits of $\tau$ to which our observations are sensitive from the limiting flux of the observations for
a given black hole mass.  Assuming $z=0.19$, we can use the theoretical decay of flare accretion as described by 
\cite{LNM02} to solve for

\begin{equation}
\tau\sim(t-t_D)\sim\left(\frac{F_{-13}}{1.5}\right)^{-3/5}\left(\frac{\Mbh}{10^7\;\Msun}\right)^{3/5} yr,
\end{equation}
where $F_{-13}$ is the limiting flux of the observation in terms of $10^{-13}$ \ecmss.

Due to the diffuse X-ray emission of the cluster, the limiting flux is dependent on the distance from the cluster center, where the number of galaxies per unit solid angle is greatest.  We must thus consider the impact of the ICM in our estimation of $N_{gal}$ and $\tau$.  To calculate our sensitivity, we therefore exclusively use values for the S band (0.3--2.5~keV).  This selection minimizes the hard contribution to the background and we expect few or no flare photons at higher energies, whereas the ICM contribution at higher energies is likely to be significant.  To determine a comprehensive limiting flux, we select a minimum of
$L_X$(0.3--2.5 keV)$=10^{42}$ erg s$^{-1}$ for a tidal flare high state to differentiate from Ultra-Luminous X-ray
sources  \cite[ULXs; e.g.,][]{Heil2009} and other similarly luminous sources, and assume $kT_{bb}\sim0.12$
keV as observed.  This  selection criterion also sets a strong lower limit to \Mbh\ for any flares in our sample, given $L_{Edd}\simgreat10^{42}$ \es\ for $\Mbh\simgreat10^4\ \Msun$.
The exposure at epoch 2006.18 is sufficiently deep that it is sensitive over the whole field to disruption events
beginning as early as the 2001.02 epoch under these selection criteria from flares similar to the SDSS J131122.15-012345.6 flare such that $\sim3\sigma$ from $\Mbh\geq10^6\;\Msun$, except in the central $\sim1\arcmin$ of the cluster.  There, X-ray emission is dominated by the ICM.

Observations at earlier \cha\ epochs (2000.29, 2001.02) are sensitive
across the field to $L_X(0.3--2.5$ keV$)\sim5\times10^{42}$ erg s$^{-1}$
at $3\sigma$.  Outside of the central $1\arcmin$ radius, early
\cha\ epochs are thus sensitive to disruptions from
$\Mbh\sim10^{6}\;\Msun$ over time scales of $\sim1.5$ yr.  The single
\xmm\ epoch at  2001.98 has a broadband $(0.1-12.0\ $keV$)\ 3\sigma$ sensitivity
$\lesssim 10^{42}$ erg s$^{-1}$ across the field.

We can estimate $N_{gal}$ by modeling the number of galaxies observed
projected in ACIS-I according to previously determined cluster
characteristics.  \cite{Lemze09} found A1689 to have a limiting radius
of 2.1 h$^{-1}$ Mpc assuming z=0.183 and a maximum radial velocity
amplitude of $\sim\vert4000\vert$ km\ s$^{-1}$ at $300 h^{-1}\ $kpc.
We use the surface number density as a function of radius that they
determined, namely $\Sigma_n=\Sigma_0/[1+(r/r_c)^2]^p+C$ where
$\Sigma_0\sim\ 1300\ $galaxies$\ h^{-2}\ $Mpc$^{-2}$,
$r_c\sim395\ h^{-1}\ $kpc, $p\sim1$ and $C$ is the number of
background galaxies. We estimate that a given \cha\ field of view with all four ACIS-I chips active plus
ACIS-S6 contains $N_{gal}\sim2100$ cluster members to a depth
of $I_{AB}=26.5$, eight magnitudes fainter than the characteristic bend at
luminosity $L^\star$ in the cluster luminosity function \citep[e.g.][]{Paolillo01}, and only
$\sim0.1N_{gal}$ will be found in the central $1\arcmin$ of bright ICM
emission.  Since fewer than $\sim10 \%$ of the galaxies covered by our observations are close enough to the cluster core to be effectively obscured by the ICM, the bright diffuse background from this region will therefore have only minor effects on the overall estimated rate and hence can be ignored.

If, from the \Mbh--L relation, we assume our candidate flare represents a characteristic
$\Mbh\sim3\times 10^{6}\;\Msun$, then the $\sim7$ years of X-ray observations for A1689 imply we are
sensitive to tidal flares of similar luminosity and duration for at least $\tau\sim8$ years. 
Combined with $N_{gal}\sim2100$, this yields $\gamma\sim6\times10^{-5}$ galaxy$^{-1}$ yr$^{-1}$ for all galaxies in the
sample.  Without taking effects of the galaxy population composition into account, this is in
excess of $\sim10^{-5}$ as found by \cite{Donley02}, within the bounds set by \cite{Luo08}, and a
factor of $\sim4$ less than estimated by \cite{Esquej08}.  \cite{Esquej08} used the \cite{FF05}
\Mbh-$\sigma$ relationship and the \cite{Wang04} rate predictions  such that
$\gamma\sim7.0~(\Mbh/10^6 \Msun)^{-0.28}\times 10^{-4}$ galaxy$^{-1}$ yr$^{-1}$.  For an assumed
$\Mbh\sim2.88\times 10^6\;\Msun$, the same formulation predicts $\gamma=5.2\times10^{-4}$ galaxy$^{-1}$ yr$^{-1}$, a factor of 9 greater than
the rate we calculated above for A1689.


However, since we expect that lower values of \Mbh\ will have markedly
shorter durations based on strength and rate of decay, suppose that for $\Mbh\sim3\times10^{6}\;\Msun$, the observable rate will likely be
dominated by black holes of \Mbh\ within an order of magnitude.  We can estimate the contributing fraction of the
cluster population by integrating a \cite{Schechter76} luminosity
function.  We use $K_s=17.02$ as above \citep{Stanford02} and a
$K_s$-band function as determined by \cite{King02}, such that
$\alpha=-1.01$ and $m_\star=14.6$.  \cite{Marconi03} determine an
$\Mbh-L_K$ relationship such that log$_{10}\Mbh \propto -2.2K$ and
there is little variation in $\Mbh-L$ relationships across several
near-infrared bands.  We integrate the Schechter function over
$K_s\pm2.2$, and then compare that value to the full galaxy population which is integrated down to $m_\star+8$ (analogously to the $I_{AB}$
limiting sensitivity determined by \cite{Lemze09}), or $M_\star+8=-16.8$ with K-corrections from \cite{King02}.  We thus find $\sim49\%$
of the galaxies to be likely contributors to the production of flares detectable by our observations.  If, therefore,
$N_{gal}\approx0.49\times2100\approx1000$, the disruption rate we infer from a {\it single} event is $1.2\times10^{-4}$.  This rate is within a
factor of two of the \cite{Esquej08} rate that makes no corrections for nondetections of flares in heavily absorbed edge-on spirals.  

But suppose that tidal flares account for some significant fraction of the other sources we detected that both displayed significant X-ray variability and could not be unambiguously eliminated from
consideration.  The highest upper bound for the flare rate would include all 14 objects as potential tidal flares.  This bound is too high as we can exclude all of the SDSS-matched objects as likely AGNs or stars based
upon the optical properties previously described.  According to \cite{Ivezic02}, SDSS morphological classification is robust to $95\%$ as faint as $r\sim 21.5$, suggesting the matched stellar objects (all of
comparable magnitude or brighter) are unlikely to be misclassified galaxies.  And a tidal optical flare of the type predicted by \cite{SQ09} ($\nu L_{\nu}\sim 10^{41}$ \es\ in $g$) is too faint to be detected
by SDSS at $z\sim 0.19$, so that such a flare could not explain starlike unresolved optical emission from a cluster member.  

Removing these 7 SDSS-matched objects from our sample, we are left with 7 remaining unmatched variable sources that lack optical identifications and are outside the FOV of all HST observations of A1689.  These objects could be background AGN that are optically fainter than the SDSS $ugriz$ limiting magnitudes.  We will now calculate upper limits to the flare rate as determined from two alternate cases that do not ignore these unmatched X-ray sources.  In the first case, we assume that all of these variable sources are cluster members.  In the second case we attempt to estimate what fraction of these unmatched sources are line-of-sight objects rather than cluster members.  But in either of these cases, any host galaxy within the cluster would have to be a dwarf galaxy.  With the \cite{Lupton05} magnitude conversions and the \cite{Lauer07} \Mbh--$\sigma$ relation, the SDSS $g$ and $r$ limiting magnitudes suggest dwarf host galaxies are unlikely to have
$\Mbh>10^{5.5}\;\Msun$.  A flare without an obvious host could also arise from an intracluster MBH ejected during an MBH merger \citep{KM08}.

In the first case, under the unlikely assumption that all optically unidentified sources were indeed tidal disruption flares from faint (dwarf) cluster galaxies, the rate for all galaxies would be $\gamma\sim 8.2\times 10^{-4}$.  But if the typical duration of a flare becomes shorter with decreasing \Mbh, the effective value of $N_{gal}\tau$ will also be less
than if we had assumed uniform flare duration.  Suppose we therefore assume a \cite{Schechter76} profile and $\tau\propto M^{3/5}$ as before, with a power law spectrum $\Gamma=2$ (using a harder spectral profile like a typical AGN and consistent with the uniform hardness ratio distribution of these objects).  Rounding our limit of $\Mbh\lesssim10^{5.5}\;\Msun$ up to $10^6\;\Msun$ for dwarf galaxies and given the Eddington-limited lower bound of $\Mbh\simgreat10^4\;\Msun$, the integrated $N_{gal}\tau$ for $10^6\;\Msun>\Mbh>10^4\;\Msun$ in our dataset is only $\sim 1400$ galaxy-years.   Thus, in this case $\gamma\sim 5.0\times10^{-3}$ for the subset of galaxies where $\Mbh\lesssim10^6\;\Msun$.  This specific rate is several times greater than even the most optimistic values (e.g. $\gamma=1.2\times 10^{-3}$ for $\Mbh=10^5\;\Msun$ from \cite{Wang04}).

In the second case, we calculate a more plausible disruption rate by using the \cite{Lemze09} galaxy distribution function to determine the likelihood
of any given galaxy being a cluster member rather than a field galaxy.  This probability decreases at greater radii from the cluster center.  Starting at $r>800$ kpc from the cluster center, even the most central of these 7 optically unidentified sources would be no more than $\sim40\%$ likely to be a cluster member.  If we consider only
the respective projected distances from the cluster center for a set of unknown galaxies, there is a $77\%$ chance that at least one of these seven galaxies is a cluster member.  The most likely or ``expectation" value is $\bar{n}=1.3$ cluster members.  Here $\bar{n}\equiv\sum_{0\leq i\leq n}p(r_i)$ where $n=7$ is the number of galaxies and $p(r_i)$ is the probability of galaxy $i$ being a cluster member at radius $r_i$ from the cluster center.  The expected number of field galaxies is therefore $n-\bar{n}=5.3$.  This estimate neglects any bias introduced in a dataset selected for X-ray emission and variability.  

To determine an upper bound for ${N_t}$ in this calculation, we consider $\bar{n}\sim 1.3$ optically unidentified X-ray-variable cluster members, plus a tidal flare in SDSS J131122.15-012345.6 and corresponding statistical error of 2.4 based on \cite{Gehrels86}.  Therefore ${N_t}\lesssim 4.7$ and $\gamma\lesssim 4.8\times 10^{-4}$.  These assumptions also indicate $\gamma\lesssim 2.6\times 10^{-3}$ for $10^6\;\Msun>\Mbh>10^4\;\Msun$, which is consistent with the predictions of \cite{Wang04}.  Thus, our observations do not exclude the
possibility that most dwarf galaxies contain MBHs in this mass range.

\section{Summary and Conclusions}

We report an X-ray flare that occurred within a projected distance of 1.6 kpc from the center of the spiral galaxy \object{SDSS J131122.15-012345.6}.  The projected positional uncertainty is 4.7 kpc.  The peak observed luminosity
($L_X$(0.3--3 keV)~$\sim5\times 10^{42}\ $erg\ s$^{-1}$, assuming it originated from a member of A1689), strong variability ($\times30$), supersoft
spectrum ($kT_{bb}=0.12$ keV), slow evolution ($\times9$ decay over $\sim2$ years), and absence of strong optical emission lines from the presumed host galaxy $\sim5$ years after the outburst are consistent with other existing examples of X-ray flares that have been classified as arising from tidal disruption events.  Although we cannot completely rule out another origin for the X-ray flare, we conclude that the most likely explanation is a flare due to the tidal disruption of a star by an MBH.  To the best of our knowledge, this is the first reported \cha-selected discovery of such a flare. 

Based on observations of a cluster with a well-measured galaxy population and over a well-defined period of time, we have therefore calculated an event rate of $\sim10^{-4}$ tidal disruption events galaxy$^{-1}$ yr$^{-1}$ under the assumption that all inactive, non-dwarf galaxies have central black holes of $\Mbh=10^6-10^8\;\Msun$ and will produce tidal flares.  The calculated event rate is consistent with previously reported theoretical and observational work.  Our results coupled with those of \cite{Cappelluti09} present a strong case for cluster surveys to detect X-ray flares from tidal disruption events.   Our work is consistent with the hypothesis that all normal galaxies (such as SDSS J131122.15-012345.6) host massive black holes at their center, but our results do not exclude other models of galaxy formation in which MBHs are rarer or (alternatively) more massive than the $\Mbh-\sigma$ relation implies \citep[e.g.][]{Volonteri05}.  

Determining the rate of tidal flares from lower mass black holes, if they occur at all, has the potential to constrain existing models of dwarf galaxy formation and evolution.  A more accurate disruption rate for normal galaxies in clusters compared to rates derived from the field could also be related to galaxy evolution in rich clusters of galaxies.  In particular, it could have implications for how galaxy mergers affect black hole mass distribution and stellar populations in galaxy clusters.
 
Flares from low mass ($\Mbh\lesssim10^6\;\Msun$) black holes may
\citep{Wang04} or may not \citep{Merritt09} contribute disproportionately to the overall flaring rate.
But if low mass black holes produce tidal flares of shorter observable duration, existing cluster X-ray observations of a typical $\sim2$ year cadence could miss tidal flares even for an actual rate that is perhaps a factor of 10 higher.  Further dedicated X-ray observations of clusters using \cha, \xmm\ or future soft-sensitive ($0.1-3.0\ $keV) missions with comparable or better
fields of view such as {\it WFXT} \citep{Murray2008}, {\it IXO} \citep{Parmar09} or {\it eRosita} \citep{eRosita07} are therefore necessary, to find events from less massive central black holes (to the extent that such less massive black holes exist) and to refine the disruption rate per galaxy.

\acknowledgments

PM and MU acknowledge the support of a NASA ADP grant NNX08AJ35G.  PM
was also supported in part by the NASA Illinois Space Consortium grant
2005-3386-02 $\backslash\backslash$ NNG05GE81H and GAANN fellowship
grant P200A060082.  We thank the referee for many very helpful comments.
We gratefully acknowledge the support of NASA for continued operations of
$HST$, \cha\ and their data archives, the ESA's support of \xmm\ and
its archives and the continued efforts of the Sloan Digital Sky
Survey.  We are grateful to Dirk Grupe for many useful discussions, and to Vincent LeBrun for fitting the HET spectrum with {\it EZ}.  PM thanks Arnold Rots for his help with the $GLvary$ code and
Olivier Ilbert for his help with $LePhare$.  MU thanks Andrew Ulmer
and Marc Freitag for helpful discussions of tidal flares.  The
Hobby-Eberly Telescope (HET) is a joint project of the University of
Texas at Austin, the Pennsylvania State University,
Ludwig-Maximillians-Universit\"{a}t M\"{u}nchen, and
Georg-August-Universit\"{a}t G\"{o}ttingen.  The HET is named in honor
of its principal benefactors, William P. Hobby and Robert E. Eberly.
The Marcario Low-Resolution Spectrograph (LRS) is a joint project of
the Hobby-Eberly Telescope partnership and the Instituto de
Astronom\'{i}a de la Universidad Nacional Aut\'{o}noma de M\'{e}xico.  This research has made use of the NASA/IPAC Extragalactic Database (NED) which is operated by the Jet Propulsion Laboratory, California Institute of Technology, under contract with the National Aeronautics and Space Administration.



{\it Facilities:} \facility{CXO(ACIS)}, \facility{XMM}, \facility{HET}, \facility{SDSS}, \facility{HST(WF/PC2)}.

\bibliographystyle{apj}  

\bibliography{apj-jour,pete_tidal,biblio_mel_marc}

\begin{thebibliography}{124}
\expandafter\ifx\csname natexlab\endcsname\relax\def\natexlab#1{#1}\fi

\bibitem[{{Adelman-McCarthy} {et~al.}(2008){Adelman-McCarthy}, {Ag{\"u}eros},
  {Allam}, {Allende Prieto}, {Anderson}, {Anderson}, {Annis}, {Bahcall},
  {Bailer-Jones}, {Baldry}, {Barentine}, {Bassett}, {Becker}, {Beers}, {Bell},
  {Berlind}, {Bernardi}, {Blanton}, {Bochanski}, {Boroski}, {Brinchmann},
  {Brinkmann}, {Brunner}, {Budav{\'a}ri}, {Carliles}, {Carr}, {Castander},
  {Cinabro}, {Cool}, {Covey}, {Csabai}, {Cunha}, {Davenport}, {Dilday}, {Doi},
  {Eisenstein}, {Evans}, {Fan}, {Finkbeiner}, {Friedman}, {Frieman},
  {Fukugita}, {G{\"a}nsicke}, {Gates}, {Gillespie}, {Glazebrook}, {Gray},
  {Grebel}, {Gunn}, {Gurbani}, {Hall}, {Harding}, {Harvanek}, {Hawley},
  {Hayes}, {Heckman}, {Hendry}, {Hindsley}, {Hirata}, {Hogan}, {Hogg}, {Hyde},
  {Ichikawa}, {Ivezi{\'c}}, {Jester}, {Johnson}, {Jorgensen}, {Juri{\'c}},
  {Kent}, {Kessler}, {Kleinman}, {Knapp}, {Kron}, {Krzesinski}, {Kuropatkin},
  {Lamb}, {Lampeitl}, {Lebedeva}, {Lee}, {Leger}, {L{\'e}pine}, {Lima}, {Lin},
  {Long}, {Loomis}, {Loveday}, {Lupton}, {Malanushenko}, {Malanushenko},
  {Mandelbaum}, {Margon}, {Marriner}, {Mart{\'{\i}}nez-Delgado}, {Matsubara},
  {McGehee}, {McKay}, {Meiksin}, {Morrison}, {Munn}, {Nakajima}, {Neilsen},
  {Newberg}, {Nichol}, {Nicinski}, {Nieto-Santisteban}, {Nitta}, {Okamura},
  {Owen}, {Oyaizu}, {Padmanabhan}, {Pan}, {Park}, {Peoples}, {Pier}, {Pope},
  {Purger}, {Raddick}, {Re Fiorentin}, {Richards}, {Richmond}, {Riess}, {Rix},
  {Rockosi}, {Sako}, {Schlegel}, {Schneider}, {Schreiber}, {Schwope}, {Seljak},
  {Sesar}, {Sheldon}, {Shimasaku}, {Sivarani}, {Smith}, {Snedden}, {Steinmetz},
  {Strauss}, {SubbaRao}, {Suto}, {Szalay}, {Szapudi}, {Szkody}, {Tegmark},
  {Thakar}, {Tremonti}, {Tucker}, {Uomoto}, {Vanden Berk}, {Vandenberg},
  {Vidrih}, {Vogeley}, {Voges}, {Vogt}, {Wadadekar}, {Weinberg}, {West},
  {White}, {Wilhite}, {Yanny}, {Yocum}, {York}, {Zehavi}, \&
  {Zucker}}]{SDSSdr6}
{Adelman-McCarthy}, J.~K., {et~al.} 2008, \apjs, 175, 297

\bibitem[{{Alexander}(2005)}]{Alexander05}
{Alexander}, T. 2005, Physics Reports, 419, 65

\bibitem[{{Alexander} \& {Kumar}(2001)}]{Alexander01}
{Alexander}, T., \& {Kumar}, P. 2001, \apj, 549, 948

\bibitem[{{Amaro-Seoane} {et~al.}(2007){Amaro-Seoane}, {Gair}, {Freitag},
  {Miller}, {Mandel}, {Cutler}, \& {Babak}}]{ASetal07}
{Amaro-Seoane}, P., {Gair}, J.~R., {Freitag}, M., {Miller}, M.~C., {Mandel},
  I., {Cutler}, C.~J., \& {Babak}, S. 2007, Classical and Quantum Gravity, 24,
  113

\bibitem[{{Andersson} \& {Madejski}(2004)}]{AM04}
{Andersson}, K.~E., \& {Madejski}, G.~M. 2004, \apj, 607, 190

\bibitem[{{Arnaud}(1996)}]{Arnaud96}
{Arnaud}, K.~A. 1996, in Astronomical Society of the Pacific Conference Series,
  Vol. 101, Astronomical Data Analysis Software and Systems V, ed.
  {G.~H.~Jacoby \& J.~Barnes}, 17--20

\bibitem[{{Ayal} {et~al.}(2000){Ayal}, {Livio}, \& {Piran}}]{ALP00}
{Ayal}, S., {Livio}, M., \& {Piran}, T. 2000, ApJ, 545, 772

\bibitem[{{Bade} {et~al.}(1996){Bade}, {Komossa}, \& {Dahlem}}]{BKD96}
{Bade}, N., {Komossa}, S., \& {Dahlem}, M. 1996, \aap, 309, L35

\bibitem[{{Barth}(2001)}]{Barth01}
{Barth}, A.~J. 2001, in Astronomical Society of the Pacific Conference Series,
  Vol. 238, Astronomical Data Analysis Software and Systems X, ed. F.~R.
  {Harnden}, Jr., F.~A. {Primini}, \& H.~E. {Payne}, 385--387

\bibitem[{{Beckmann} {et~al.}(2006){Beckmann}, {Gehrels}, {Shrader}, \&
  {Soldi}}]{Beckmann06}
{Beckmann}, V., {Gehrels}, N., {Shrader}, C.~R., \& {Soldi}, S. 2006, \apj,
  638, 642

\bibitem[{{Blanton} {et~al.}(2003){Blanton}, {Hogg}, {Bahcall}, {Baldry},
  {Brinkmann}, {Csabai}, {Eisenstein}, {Fukugita}, {Gunn}, {Ivezi{\'c}},
  {Lamb}, {Lupton}, {Loveday}, {Munn}, {Nichol}, {Okamura}, {Schlegel},
  {Shimasaku}, {Strauss}, {Vogeley}, \& {Weinberg}}]{Blanton03}
{Blanton}, M.~R., {et~al.} 2003, \apj, 594, 186

\bibitem[{{Bogdanovi{\' c}} {et~al.}(2004){Bogdanovi{\' c}}, {Eracleous},
  {Mahadevan}, {Sigurdsson}, \& {Laguna}}]{BogdanovicEtAl04}
{Bogdanovi{\' c}}, T., {Eracleous}, M., {Mahadevan}, S., {Sigurdsson}, S., \&
  {Laguna}, P. 2004, ApJ, 610, 707

\bibitem[{{Boller} {et~al.}(1997){Boller}, {Brandt}, {Fabian}, \&
  {Fink}}]{Boller97}
{Boller}, T., {Brandt}, W.~N., {Fabian}, A.~C., \& {Fink}, H.~H. 1997, \mnras,
  289, 393

\bibitem[{{Boller} {et~al.}(1993){Boller}, {Truemper}, {Molendi}, {Fink},
  {Schaeidt}, {Caulet}, \& {Dennefeld}}]{Boller93}
{Boller}, T., {Truemper}, J., {Molendi}, S., {Fink}, H., {Schaeidt}, S.,
  {Caulet}, A., \& {Dennefeld}, M. 1993, \aap, 279, 53

\bibitem[{{Brandt} {et~al.}(1995){Brandt}, {Pounds}, \& {Fink}}]{BPF95}
{Brandt}, W.~N., {Pounds}, K.~A., \& {Fink}, H. 1995, \mnras, 273, L47

\bibitem[{{Brassart} \& {Luminet}(2008)}]{BL08}
{Brassart}, M., \& {Luminet}, J. 2008, \aap, 481, 259

\bibitem[{{Campana} {et~al.}(2006){Campana}, {Mangano}, {Blustin}, {Brown},
  {Burrows}, {Chincarini}, {Cummings}, {Cusumano}, {Della Valle}, {Malesani},
  {M{\'e}sz{\'a}ros}, {Nousek}, {Page}, {Sakamoto}, {Waxman}, {Zhang}, {Dai},
  {Gehrels}, {Immler}, {Marshall}, {Mason}, {Moretti}, {O'Brien}, {Osborne},
  {Page}, {Romano}, {Roming}, {Tagliaferri}, {Cominsky}, {Giommi}, {Godet},
  {Kennea}, {Krimm}, {Angelini}, {Barthelmy}, {Boyd}, {Palmer}, {Wells}, \&
  {White}}]{Campana06}
{Campana}, S., {et~al.} 2006, \nat, 442, 1008

\bibitem[{{Cappelluti} {et~al.}(2009){Cappelluti}, {Ajello}, {Rebusco},
  {Komossa}, {Bongiorno}, {Clemens}, {Salvato}, {Esquej}, {Aldcroft},
  {Greiner}, \& {Quintana}}]{Cappelluti09}
{Cappelluti}, N., {et~al.} 2009, \aap, 495, L9

\bibitem[{{Carter} \& {Luminet}(1982)}]{CL82}
{Carter}, B., \& {Luminet}, J.~P. 1982, \nat, 296, 211

\bibitem[{{Churazov} {et~al.}(1996){Churazov}, {Gilfanov}, {Forman}, \&
  {Jones}}]{chur96}
{Churazov}, E., {Gilfanov}, M., {Forman}, W., \& {Jones}, C. 1996, \apj, 471,
  673

\bibitem[{{Collister} \& {Lahav}(2004)}]{ANNz}
{Collister}, A.~A., \& {Lahav}, O. 2004, \pasp, 116, 345

\bibitem[{{Dickey} \& {Lockman}(1990)}]{Dickey90}
{Dickey}, J.~M., \& {Lockman}, F.~J. 1990, \araa, 28, 215

\bibitem[{{Do} {et~al.}(2009){Do}, {Ghez}, {Morris}, {Lu}, {Matthews}, {Yelda},
  \& {Larkin}}]{Do09}
{Do}, T., {Ghez}, A.~M., {Morris}, M.~R., {Lu}, J.~R., {Matthews}, K., {Yelda},
  S., \& {Larkin}, J. 2009, \apj, 703, 1323

\bibitem[{{Donley} {et~al.}(2002){Donley}, {Brandt}, {Eracleous}, \&
  {Boller}}]{Donley02}
{Donley}, J.~L., {Brandt}, W.~N., {Eracleous}, M., \& {Boller}, T. 2002, \aj,
  124, 1308

\bibitem[{{Dorman} \& {Arnaud}(2001)}]{Dorman01}
{Dorman}, B., \& {Arnaud}, K.~A. 2001, in Astronomical Society of the Pacific
  Conference Series, Vol. 238, Astronomical Data Analysis Software and Systems
  X, ed. {F.~R.~Harnden Jr., F.~A.~Primini, \& H.~E.~Payne}, 415--418

\bibitem[{{Duc} {et~al.}(2002){Duc}, {Poggianti}, {Fadda}, {Elbaz}, {Flores},
  {Chanial}, {Franceschini}, {Moorwood}, \& {Cesarsky}}]{Duc02}
{Duc}, P.-A., {et~al.} 2002, \aap, 382, 60

\bibitem[{{Elvis} {et~al.}(1994){Elvis}, {Wilkes}, {McDowell}, {Green},
  {Bechtold}, {Willner}, {Oey}, {Polomski}, \& {Cutri}}]{Elvis94}
{Elvis}, M., {et~al.} 1994, \apjs, 95, 1

\bibitem[{{Esquej} {et~al.}(2007){Esquej}, {Saxton}, {Freyberg}, {Read},
  {Altieri}, {Sanchez-Portal}, \& {Hasinger}}]{Esquej07}
{Esquej}, P., {Saxton}, R.~D., {Freyberg}, M.~J., {Read}, A.~M., {Altieri}, B.,
  {Sanchez-Portal}, M., \& {Hasinger}, G. 2007, \aap, 462, L49

\bibitem[{{Esquej} {et~al.}(2008){Esquej}, {Saxton}, {Komossa}, {Read},
  {Freyberg}, {Hasinger}, {Garc{\'{\i}}a-Hern{\'a}ndez}, {Lu}, {Zaur{\'{\i}}n},
  {S{\'a}nchez-Portal}, \& {Zhou}}]{Esquej08}
{Esquej}, P., {et~al.} 2008, \aap, 489, 543

\bibitem[{{Estrada} {et~al.}(2007){Estrada}, {Annis}, {Diehl}, {Hall}, {Las},
  {Lin}, {Makler}, {Merritt}, {Scarpine}, {Allam}, \& {Tucker}}]{Estrada07}
{Estrada}, J., {et~al.} 2007, \apj, 660, 1176

\bibitem[{{Evans} \& {Kochanek}(1989)}]{EK89}
{Evans}, C.~R., \& {Kochanek}, C.~S. 1989, ApJ Lett., 346, L13

\bibitem[{{Feigelson} {et~al.}(2002){Feigelson}, {Broos}, {Gaffney}, {Garmire},
  {Hillenbrand}, {Pravdo}, {Townsley}, \& {Tsuboi}}]{Feigelson02}
{Feigelson}, E.~D., {Broos}, P., {Gaffney}, III, J.~A., {Garmire}, G.,
  {Hillenbrand}, L.~A., {Pravdo}, S.~H., {Townsley}, L., \& {Tsuboi}, Y. 2002,
  \apj, 574, 258

\bibitem[{{Ferrarese} \& {Ford}(2005)}]{FF05}
{Ferrarese}, L., \& {Ford}, H. 2005, Space Science Reviews, 116, 523

\bibitem[{{Ferrarese} {et~al.}(2006){Ferrarese}, {C{\^o}t{\'e}}, {Dalla
  Bont{\`a}}, {Peng}, {Merritt}, {Jord{\'a}n}, {Blakeslee}, {Ha{\c s}egan},
  {Mei}, {Piatek}, {Tonry}, \& {West}}]{FerrareseEtAl06}
{Ferrarese}, L., {et~al.} 2006, ApJ, 644, L21

\bibitem[{{Fruscione} {et~al.}(2006){Fruscione}, {McDowell}, {Allen},
  {Brickhouse}, {Burke}, {Davis}, {Durham}, {Elvis}, {Galle}, {Harris},
  {Huenemoerder}, {Houck}, {Ishibashi}, {Karovska}, {Nicastro}, {Noble},
  {Nowak}, {Primini}, {Siemiginowska}, {Smith}, \& {Wise}}]{ciao06}
{Fruscione}, A., {et~al.} 2006, in Society of Photo-Optical Instrumentation
  Engineers (SPIE) Conference Series, Vol. 6270, Society of Photo-Optical
  Instrumentation Engineers (SPIE) Conference Series

\bibitem[{{Gehrels}(1986)}]{Gehrels86}
{Gehrels}, N. 1986, \apj, 303, 336

\bibitem[{{Genzel} {et~al.}(2003){Genzel}, {Sch{\"o}del}, {Ott}, {Eisenhauer},
  {Hofmann}, {Lehnert}, {Eckart}, {Alexander}, {Sternberg}, {Lenzen},
  {Cl{\'e}net}, {Lacombe}, {Rouan}, {Renzini}, \& {Tacconi-Garman}}]{Genzel03}
{Genzel}, R., {et~al.} 2003, \apj, 594, 812

\bibitem[{{Gezari} {et~al.}(2003){Gezari}, {Halpern}, {Komossa}, {Grupe}, \&
  {Leighly}}]{Gezari03}
{Gezari}, S., {Halpern}, J.~P., {Komossa}, S., {Grupe}, D., \& {Leighly}, K.~M.
  2003, \apj, 592, 42

\bibitem[{{Gezari} {et~al.}(2006){Gezari}, {Martin}, {Milliard}, {Basa},
  {Halpern}, {Forster}, {Friedman}, {Morrissey}, {Neff}, {Schiminovich},
  {Seibert}, {Small}, \& {Wyder}}]{Gezari06}
{Gezari}, S., {et~al.} 2006, \apjl, 653, L25

\bibitem[{{Gezari} {et~al.}(2008{\natexlab{a}}){Gezari}, {Dessart}, {Basa},
  {Martin}, {Neill}, {Woosley}, {Hillier}, {Bazin}, {Forster}, {Friedman}, {Le
  Du}, {Mazure}, {Morrissey}, {Neff}, {Schiminovich}, \& {Wyder}}]{GezariSN08}
---. 2008{\natexlab{a}}, \apjl, 683, L131

\bibitem[{{Gezari} {et~al.}(2008{\natexlab{b}}){Gezari}, {Basa}, {Martin},
  {Bazin}, {Forster}, {Milliard}, {Halpern}, {Friedman}, {Morrissey}, {Neff},
  {Schiminovich}, {Seibert}, {Small}, \& {Wyder}}]{Gezari08}
---. 2008{\natexlab{b}}, \apj, 676, 944

\bibitem[{{Gezari} {et~al.}(2009){Gezari}, {Heckman}, {Cenko}, {Eracleous},
  {Forster}, {Gon{\c c}alves}, {Martin}, {Morrissey}, {Neff}, {Seibert},
  {Schiminovich}, \& {Wyder}}]{Gezari09}
---. 2009, \apj, 698, 1367

\bibitem[{{Ghez} {et~al.}(2003){Ghez}, {Becklin}, {Duchjne}, {Hornstein},
  {Morris}, {Salim}, \& {Tanner}}]{Ghez03}
{Ghez}, A.~M., {Becklin}, E., {Duchjne}, G., {Hornstein}, S., {Morris}, M.,
  {Salim}, S., \& {Tanner}, A. 2003, Astronomische Nachrichten Supplement, 324,
  527

\bibitem[{{Giacconi} {et~al.}(2001){Giacconi}, {Rosati}, {Tozzi}, {Nonino},
  {Hasinger}, {Norman}, {Bergeron}, {Borgani}, {Gilli}, {Gilmozzi}, \&
  {Zheng}}]{Giacconi01}
{Giacconi}, R., {et~al.} 2001, \apj, 551, 624

\bibitem[{{Graham} \& {Driver}(2005)}]{GD05}
{Graham}, A.~W., \& {Driver}, S.~P. 2005, Publications of the Astronomical
  Society of Australia, 22, 118

\bibitem[{{Greene} {et~al.}(2008){Greene}, {Ho}, \& {Barth}}]{GHB08}
{Greene}, J.~E., {Ho}, L.~C., \& {Barth}, A.~J. 2008, \apj, 688, 159

\bibitem[{{Gregory} \& {Loredo}(1992)}]{GL92}
{Gregory}, P.~C., \& {Loredo}, T.~J. 1992, \apj, 398, 146

\bibitem[{{Grupe}(2004)}]{Grupe04}
{Grupe}, D. 2004, \aj, 127, 1799

\bibitem[{{Grupe} {et~al.}(1995{\natexlab{a}}){Grupe}, {Beuerman}, {Mannheim},
  {Thomas}, {Fink}, \& {de Martino}}]{Grupe95a}
{Grupe}, D., {Beuerman}, K., {Mannheim}, K., {Thomas}, H., {Fink}, H.~H., \&
  {de Martino}, D. 1995{\natexlab{a}}, \aap, 300, L21+

\bibitem[{{Grupe} {et~al.}(1995{\natexlab{b}}){Grupe}, {Beuermann}, {Mannheim},
  {Bade}, {Thomas}, {de Martino}, \& {Schwope}}]{Grupe95b}
{Grupe}, D., {Beuermann}, K., {Mannheim}, K., {Bade}, N., {Thomas}, H., {de
  Martino}, D., \& {Schwope}, A. 1995{\natexlab{b}}, \aap, 299, L5+

\bibitem[{{Grupe} {et~al.}(2008{\natexlab{a}}){Grupe}, {Komossa}, {Gallo},
  {Fabian}, {Larsson}, {Pradhan}, {Xu}, \& {Miniutti}}]{Grupe08}
{Grupe}, D., {Komossa}, S., {Gallo}, L.~C., {Fabian}, A.~C., {Larsson}, J.,
  {Pradhan}, A.~K., {Xu}, D., \& {Miniutti}, G. 2008{\natexlab{a}}, \apj, 681,
  982

\bibitem[{{Grupe} {et~al.}(2008{\natexlab{b}}){Grupe}, {Leighly}, \&
  {Komossa}}]{GLK08}
{Grupe}, D., {Leighly}, K.~M., \& {Komossa}, S. 2008{\natexlab{b}}, \aj, 136,
  2343

\bibitem[{{Grupe} {et~al.}(1999){Grupe}, {Thomas}, \& {Leighly}}]{GTL99}
{Grupe}, D., {Thomas}, H., \& {Leighly}, K.~M. 1999, \aap, 350, L31

\bibitem[{{Grupe} {et~al.}(2004){Grupe}, {Wills}, {Leighly}, \&
  {Meusinger}}]{Grupeetal04}
{Grupe}, D., {Wills}, B.~J., {Leighly}, K.~M., \& {Meusinger}, H. 2004, \aj,
  127, 156

\bibitem[{{Grupe} {et~al.}(2010){Grupe}, {Burrows}, {Wu}, {Wang}, {Zhang},
  {Liang}, {Garmire}, {Nousek}, {Gehrels}, {Ricker}, \& {Bautz}}]{Grupe10}
{Grupe}, D., {et~al.} 2010, \apj, 711, 1008

\bibitem[{{Halkola} {et~al.}(2006){Halkola}, {Seitz}, \&
  {Pannella}}]{Halkola06}
{Halkola}, A., {Seitz}, S., \& {Pannella}, M. 2006, \mnras, 372, 1425

\bibitem[{{Halpern} {et~al.}(2004){Halpern}, {Gezari}, \&
  {Komossa}}]{Halpern04}
{Halpern}, J.~P., {Gezari}, S., \& {Komossa}, S. 2004, \apj, 604, 572

\bibitem[{{Hansen} {et~al.}(2004){Hansen}, {Kawaler}, \& {Trimble}}]{HKT}
{Hansen}, C.~J., {Kawaler}, S.~D., \& {Trimble}, V. 2004, {Stellar interiors :
  physical principles, structure, and evolution}, ed. C.~J. {Hansen}, S.~D.
  {Kawaler}, \& V.~{Trimble}

\bibitem[{{Heil} {et~al.}(2009){Heil}, {Vaughan}, \& {Roberts}}]{Heil2009}
{Heil}, L.~M., {Vaughan}, S., \& {Roberts}, T.~P. 2009, \mnras, 397, 1061

\bibitem[{{Heinke} {et~al.}(2003){Heinke}, {Grindlay}, {Lugger}, {Cohn},
  {Edmonds}, {Lloyd}, \& {Cool}}]{Heinke03}
{Heinke}, C.~O., {Grindlay}, J.~E., {Lugger}, P.~M., {Cohn}, H.~N., {Edmonds},
  P.~D., {Lloyd}, D.~A., \& {Cool}, A.~M. 2003, \apj, 598, 501

\bibitem[{{Ilbert} {et~al.}(2006){Ilbert}, {Arnouts}, {McCracken},
  {Bolzonella}, {Bertin}, {Le F{\`e}vre}, {Mellier}, {Zamorani}, {Pell{\`o}},
  {Iovino}, {Tresse}, {Le Brun}, {Bottini}, {Garilli}, {Maccagni}, {Picat},
  {Scaramella}, {Scodeggio}, {Vettolani}, {Zanichelli}, {Adami}, {Bardelli},
  {Cappi}, {Charlot}, {Ciliegi}, {Contini}, {Cucciati}, {Foucaud}, {Franzetti},
  {Gavignaud}, {Guzzo}, {Marano}, {Marinoni}, {Mazure}, {Meneux}, {Merighi},
  {Paltani}, {Pollo}, {Pozzetti}, {Radovich}, {Zucca}, {Bondi}, {Bongiorno},
  {Busarello}, {de La Torre}, {Gregorini}, {Lamareille}, {Mathez}, {Merluzzi},
  {Ripepi}, {Rizzo}, \& {Vergani}}]{LePhare}
{Ilbert}, O., {et~al.} 2006, \aap, 457, 841

\bibitem[{{Immler} \& {Lewin}(2003)}]{Immler03}
{Immler}, S., \& {Lewin}, W.~H.~G. 2003, in Lecture Notes in Physics, Berlin
  Springer Verlag, Vol. 598, Supernovae and Gamma-Ray Bursters, ed.
  K.~{Weiler}, 91--111

\bibitem[{{Immler} {et~al.}(2008){Immler}, {Modjaz}, {Landsman}, {Bufano},
  {Brown}, {Milne}, {Dessart}, {Holland}, {Koss}, {Pooley}, {Kirshner},
  {Filippenko}, {Panagia}, {Chevalier}, {Mazzali}, {Gehrels}, {Petre},
  {Burrows}, {Nousek}, {Roming}, {Pian}, {Soderberg}, \& {Greiner}}]{Immler08}
{Immler}, S., {et~al.} 2008, \apjl, 674, L85

\bibitem[{{Ivezi{\'c}} {et~al.}(2002){Ivezi{\'c}}, {Menou}, {Knapp}, {Strauss},
  {Lupton}, {Vanden Berk}, {Richards}, {Tremonti}, {Weinstein}, {Anderson},
  {Bahcall}, {Becker}, {Bernardi}, {Blanton}, {Eisenstein}, {Fan},
  {Finkbeiner}, {Finlator}, {Frieman}, {Gunn}, {Hall}, {Kim}, {Kinkhabwala},
  {Narayanan}, {Rockosi}, {Schlegel}, {Schneider}, {Strateva}, {SubbaRao},
  {Thakar}, {Voges}, {White}, {Yanny}, {Brinkmann}, {Doi}, {Fukugita},
  {Hennessy}, {Munn}, {Nichol}, \& {York}}]{Ivezic02}
{Ivezi{\'c}}, {\v Z}., {et~al.} 2002, \aj, 124, 2364

\bibitem[{{Jenkins} {et~al.}(2007){Jenkins}, {Hornschemeier}, {Mobasher},
  {Alexander}, \& {Bauer}}]{Jenkins07}
{Jenkins}, L.~P., {Hornschemeier}, A.~E., {Mobasher}, B., {Alexander}, D.~M.,
  \& {Bauer}, F.~E. 2007, \apj, 666, 846

\bibitem[{{Jennrich}(2004)}]{Jennrich04}
{Jennrich}, O. 2004, in Optical Fabrication, Metrology, and Material
  Advancements for Telescopes. Edited by Atad-Ettedgui, Eli; Dierickx,
  Philippe. Proceedings of the SPIE, Volume 5500, pp. 113-119 (2004)., ed.
  J.~{Hough} \& G.~H. {Sanders}, 113--119

\bibitem[{{Khokhlov} \& {Melia}(1996)}]{KM96}
{Khokhlov}, A., \& {Melia}, F. 1996, \apjl, 457, L61

\bibitem[{{Khokhlov} {et~al.}(1993){Khokhlov}, {Novikov}, \&
  {Pethick}}]{KNP93b}
{Khokhlov}, A., {Novikov}, I.~D., \& {Pethick}, C.~J. 1993, ApJ, 418, 181

\bibitem[{{Kim} {et~al.}(2007){Kim}, {Kim}, {Wilkes}, {Green}, {Kim},
  {Anderson}, {Barkhouse}, {Evans}, {Ivezi{\'c}}, {Karovska}, {Kashyap}, {Lee},
  {Maksym}, {Mossman}, {Silverman}, \& {Tananbaum}}]{ChaMPCat}
{Kim}, M., {et~al.} 2007, \apjs, 169, 401

\bibitem[{{King} {et~al.}(2002){King}, {Clowe}, {Lidman}, {Schneider}, {Erben},
  {Kneib}, \& {Meylan}}]{King02}
{King}, L.~J., {Clowe}, D.~I., {Lidman}, C., {Schneider}, P., {Erben}, T.,
  {Kneib}, J.-P., \& {Meylan}, G. 2002, \aap, 385, L5

\bibitem[{{Kobayashi} {et~al.}(2004){Kobayashi}, {Laguna}, {Phinney}, \&
  {M{\'e}sz{\'a}ros}}]{Kobayashi04}
{Kobayashi}, S., {Laguna}, P., {Phinney}, E.~S., \& {M{\'e}sz{\'a}ros}, P.
  2004, \apj, 615, 855

\bibitem[{{Komossa}(2005)}]{Komossa05}
{Komossa}, S. 2005, in Growing Black Holes: Accretion in a Cosmological
  Context, ed. A.~{Merloni}, S.~{Nayakshin}, \& R.~A. {Sunyaev}, 159--163

\bibitem[{{Komossa} \& {Bade}(1999)}]{KB99}
{Komossa}, S., \& {Bade}, N. 1999, \aap, 343, 775

\bibitem[{{Komossa} \& {Greiner}(1999)}]{KG99}
{Komossa}, S., \& {Greiner}, J. 1999, \aap, 349, L45

\bibitem[{{Komossa} {et~al.}(2004){Komossa}, {Halpern}, {Schartel}, {Hasinger},
  {Santos-Lleo}, \& {Predehl}}]{Komossa04}
{Komossa}, S., {Halpern}, J., {Schartel}, N., {Hasinger}, G., {Santos-Lleo},
  M., \& {Predehl}, P. 2004, \apjl, 603, L17

\bibitem[{{Komossa} \& {Merritt}(2008)}]{KM08}
{Komossa}, S., \& {Merritt}, D. 2008, \apjl, 683, L21

\bibitem[{{Landsman}(1993)}]{Landsman93}
{Landsman}, W.~B. 1993, in Astronomical Society of the Pacific Conference
  Series, Vol.~52, Astronomical Data Analysis Software and Systems II, ed.
  R.~J. {Hanisch}, R.~J.~V. {Brissenden}, \& J.~{Barnes}, 246--248

\bibitem[{{Lauer} {et~al.}(2007){Lauer}, {Faber}, {Richstone}, {Gebhardt},
  {Tremaine}, {Postman}, {Dressler}, {Aller}, {Filippenko}, {Green}, {Ho},
  {Kormendy}, {Magorrian}, \& {Pinkney}}]{Lauer07}
{Lauer}, T.~R., {et~al.} 2007, \apj, 662, 808

\bibitem[{{Le F{\`e}vre} {et~al.}(2005){Le F{\`e}vre}, {Guzzo}, {Meneux},
  {Pollo}, {Cappi}, {Colombi}, {Iovino}, {Marinoni}, {McCracken}, {Scaramella},
  {Bottini}, {Garilli}, {Le Brun}, {Maccagni}, {Picat}, {Scodeggio}, {Tresse},
  {Vettolani}, {Zanichelli}, {Adami}, {Arnaboldi}, {Arnouts}, {Bardelli},
  {Blaizot}, {Bolzonella}, {Charlot}, {Ciliegi}, {Contini}, {Foucaud},
  {Franzetti}, {Gavignaud}, {Ilbert}, {Marano}, {Mathez}, {Mazure}, {Merighi},
  {Paltani}, {Pell{\`o}}, {Pozzetti}, {Radovich}, {Zamorani}, {Zucca}, {Bondi},
  {Bongiorno}, {Busarello}, {Lamareille}, {Mellier}, {Merluzzi}, {Ripepi}, \&
  {Rizzo}}]{VVDS05}
{Le F{\`e}vre}, O., {et~al.} 2005, \aap, 439, 877

\bibitem[{{Lemze} {et~al.}(2009){Lemze}, {Broadhurst}, {Rephaeli}, {Barkana},
  \& {Umetsu}}]{Lemze09}
{Lemze}, D., {Broadhurst}, T., {Rephaeli}, Y., {Barkana}, R., \& {Umetsu}, K.
  2009, \apj, 701, 1336

\bibitem[{{Li} {et~al.}(2002){Li}, {Narayan}, \& {Menou}}]{LNM02}
{Li}, L., {Narayan}, R., \& {Menou}, K. 2002, ApJ, 576, 753

\bibitem[{{Lodato} {et~al.}(2009){Lodato}, {King}, \& {Pringle}}]{Lodato09}
{Lodato}, G., {King}, A.~R., \& {Pringle}, J.~E. 2009, \mnras, 392, 332

\bibitem[{{Loeb} \& {Ulmer}(1997)}]{Loeb97}
{Loeb}, A., \& {Ulmer}, A. 1997, \apj, 489, 573

\bibitem[{{Luo} {et~al.}(2008){Luo}, {Brandt}, {Steffen}, \& {Bauer}}]{Luo08}
{Luo}, B., {Brandt}, W.~N., {Steffen}, A.~T., \& {Bauer}, F.~E. 2008, \apj,
  674, 122

\bibitem[{{Lupton}(2005)}]{Lupton05}
{Lupton}, R. 2005, {SDSS Photometric Equations}, {Batavia, IL: SDSS,
  http://www.sdss.org/ sdr6/algorithms/sdssUBVRITransform.html}

\bibitem[{{Magorrian} \& {Tremaine}(1999)}]{MT99}
{Magorrian}, J., \& {Tremaine}, S. 1999, MNRAS, 309, 447

\bibitem[{{Magorrian} {et~al.}(1998){Magorrian}, {Tremaine}, {Richstone},
  {Bender}, {Bower}, {Dressler}, {Faber}, {Gebhardt}, {Green}, {Grillmair},
  {Kormendy}, \& {Lauer}}]{Magorrian98}
{Magorrian}, J., {et~al.} 1998, \aj, 115, 2285

\bibitem[{{Marconi} \& {Hunt}(2003)}]{Marconi03}
{Marconi}, A., \& {Hunt}, L.~K. 2003, \apjl, 589, L21

\bibitem[{{Marconi} {et~al.}(2004){Marconi}, {Risaliti}, {Gilli}, {Hunt},
  {Maiolino}, \& {Salvati}}]{Marconi04}
{Marconi}, A., {Risaliti}, G., {Gilli}, R., {Hunt}, L.~K., {Maiolino}, R., \&
  {Salvati}, M. 2004, \mnras, 351, 169

\bibitem[{{Martini} {et~al.}(2007){Martini}, {Mulchaey}, \&
  {Kelson}}]{Martini07}
{Martini}, P., {Mulchaey}, J.~S., \& {Kelson}, D.~D. 2007, \apj, 664, 761

\bibitem[{{Mathur} \& {Grupe}(2005)}]{MG05}
{Mathur}, S., \& {Grupe}, D. 2005, \apj, 633, 688

\bibitem[{{McHardy} {et~al.}(2004){McHardy}, {Papadakis}, {Uttley}, {Page}, \&
  {Mason}}]{McHardy04}
{McHardy}, I.~M., {Papadakis}, I.~E., {Uttley}, P., {Page}, M.~J., \& {Mason},
  K.~O. 2004, \mnras, 348, 783

\bibitem[{{Merritt}(2009)}]{Merritt09}
{Merritt}, D. 2009, \apj, 694, 959

\bibitem[{{Milosavljevi{\'c}} {et~al.}(2006){Milosavljevi{\'c}}, {Merritt}, \&
  {Ho}}]{Milos06}
{Milosavljevi{\'c}}, M., {Merritt}, D., \& {Ho}, L.~C. 2006, \apj, 652, 120

\bibitem[{{Murray} {et~al.}(2008){Murray}, {Norman}, {Ptak}, {Giacconi},
  {Weisskopf}, {Ramsey}, {Bautz}, {Vikhliniin}, {Brandt}, {Rosati}, {Weaver},
  {Allen}, \& {Flanagan}}]{Murray2008}
{Murray}, S.~S., {et~al.} 2008, in Society of Photo-Optical Instrumentation
  Engineers (SPIE) Conference Series, Vol. 7011, Society of Photo-Optical
  Instrumentation Engineers (SPIE) Conference Series

\bibitem[{{Paolillo} {et~al.}(2001){Paolillo}, {Andreon}, {Longo}, {Puddu},
  {Gal}, {Scaramella}, {Djorgovski}, \& {de Carvalho}}]{Paolillo01}
{Paolillo}, M., {Andreon}, S., {Longo}, G., {Puddu}, E., {Gal}, R.~R.,
  {Scaramella}, R., {Djorgovski}, S.~G., \& {de Carvalho}, R. 2001, \aap, 367,
  59

\bibitem[{{Parmar}(2009)}]{Parmar09}
{Parmar}, A. 2009, in High Resolution X-ray Spectroscopy: Towards IXO,
  Proceedings of the international workshop held at the Mullard Space Science
  Laboratory of University College London, Holmbury St Mary, Dorking, Surrey,
  UK, March 19 - 20, 2009, Ed.s Branduardi-Raymont, G. and Blustin, A.,
  published electronically at http://www.mssl.ucl.ac.uk/\~{}ajb/workshop3/
  index.html, p.E32

\bibitem[{{Phinney}(1989)}]{Phinney89}
{Phinney}, E.~S. 1989, in IAU Symp. 136: The Center of the Galaxy, ed.
  M.~{Morris}, Vol. 136, 543

\bibitem[{{Predehl} {et~al.}(2007){Predehl}, {Andritschke}, {Bornemann},
  {Br{\"a}uninger}, {Briel}, {Brunner}, {Burkert}, {Dennerl}, {Eder},
  {Freyberg}, {Friedrich}, {F{\"u}rmetz}, {Hartmann}, {Hartner}, {Hasinger},
  {Herrmann}, {Holl}, {Huber}, {Kendziorra}, {Kink}, {Meidinger}, {M{\"u}ller},
  {Pavlinsky}, {Pfeffermann}, {Roh{\'e}}, {Santangelo}, {Schmitt}, {Schwope},
  {Steinmetz}, {Str{\"u}der}, {Sunyaev}, {Tiedemann}, {Vongehr}, {Wilms},
  {Erhard}, {Gutruf}, {Jugler}, {Kampf}, {Graue}, {Citterio}, {Valsecci},
  {Vernani}, \& {Zimmerman}}]{eRosita07}
{Predehl}, P., {et~al.} 2007, in Society of Photo-Optical Instrumentation
  Engineers (SPIE) Conference Series, Vol. 6686, Society of Photo-Optical
  Instrumentation Engineers (SPIE) Conference Series

\bibitem[{{Rees}(1988)}]{Rees88}
{Rees}, M.~J. 1988, Nature, 333, 523

\bibitem[{{ROSAT Consortium}(2000)}]{2RXP}
{ROSAT Consortium}. 2000, VizieR Online Data Catalog, 9030, 0

\bibitem[{{ROSAT Scientific Team}(2000)}]{1RXH}
{ROSAT Scientific Team}. 2000, VizieR Online Data Catalog, 9028, 0

\bibitem[{{Rots}(2006)}]{Rots06}
{Rots}, A.~H. 2006, in Astronomical Society of the Pacific Conference Series,
  Vol. 351, Astronomical Data Analysis Software and Systems XV, ed.
  C.~{Gabriel}, C.~{Arviset}, D.~{Ponz}, \& S.~{Enrique}, 73

\bibitem[{{Schechter}(1976)}]{Schechter76}
{Schechter}, P. 1976, \apj, 203, 297

\bibitem[{{Schlegel} \& {Petre}(2006)}]{SP06}
{Schlegel}, E.~M., \& {Petre}, R. 2006, \apj, 646, 378

\bibitem[{{S{\'e}rsic}(1963)}]{Sersic}
{S{\'e}rsic}, J.~L. 1963, Boletin de la Asociacion Argentina de Astronomia La
  Plata Argentina, 6, 41

\bibitem[{{Sesana} {et~al.}(2004){Sesana}, {Haardt}, {Madau}, \&
  {Volonteri}}]{SHMV04}
{Sesana}, A., {Haardt}, F., {Madau}, P., \& {Volonteri}, M. 2004, ApJ, 611, 623

\bibitem[{{Sesana} {et~al.}(2005){Sesana}, {Haardt}, {Madau}, \&
  {Volonteri}}]{SHMV05}
---. 2005, ApJ, 623, 23

\bibitem[{{Shapiro} \& {Teukolsky}(1986)}]{ST86}
{Shapiro}, S.~L., \& {Teukolsky}, S.~A. 1986, {Black Holes, White Dwarfs and
  Neutron Stars: The Physics of Compact Objects}, ed. {Shapiro, S.~L.~\&
  Teukolsky, S.~A.}

\bibitem[{{Sigurdsson}(2003)}]{Sigurdsson03}
{Sigurdsson}, S. 2003, Classical and Quantum Gravity, 20, 45

\bibitem[{{Soderberg} {et~al.}(2008){Soderberg}, {Berger}, {Page}, {Schady},
  {Parrent}, {Pooley}, {Wang}, {Ofek}, {Cucchiara}, {Rau}, {Waxman}, {Simon},
  {Bock}, {Milne}, {Page}, {Barentine}, {Barthelmy}, {Beardmore}, {Bietenholz},
  {Brown}, {Burrows}, {Burrows}, {Byrngelson}, {Cenko}, {Chandra}, {Cummings},
  {Fox}, {Gal-Yam}, {Gehrels}, {Immler}, {Kasliwal}, {Kong}, {Krimm},
  {Kulkarni}, {Maccarone}, {M{\'e}sz{\'a}ros}, {Nakar}, {O'Brien}, {Overzier},
  {de Pasquale}, {Racusin}, {Rea}, \& {York}}]{Soderberg08}
{Soderberg}, A.~M., {et~al.} 2008, \nat, 453, 469

\bibitem[{{Stanford} {et~al.}(2002){Stanford}, {Eisenhardt}, {Dickinson},
  {Holden}, \& {De Propris}}]{Stanford02}
{Stanford}, S.~A., {Eisenhardt}, P.~R., {Dickinson}, M., {Holden}, B.~P., \&
  {De Propris}, R. 2002, \apjs, 142, 153

\bibitem[{{Strubbe} \& {Quataert}(2009)}]{SQ09}
{Strubbe}, L.~E., \& {Quataert}, E. 2009, \mnras, 400, 2070

\bibitem[{{Struble} \& {Rood}(1999)}]{Struble99}
{Struble}, M.~F., \& {Rood}, H.~J. 1999, \apjs, 125, 35

\bibitem[{{Szokoly} {et~al.}(2004){Szokoly}, {Bergeron}, {Hasinger}, {Lehmann},
  {Kewley}, {Mainieri}, {Nonino}, {Rosati}, {Giacconi}, {Gilli}, {Gilmozzi},
  {Norman}, {Romaniello}, {Schreier}, {Tozzi}, {Wang}, {Zheng}, \&
  {Zirm}}]{Szokoly04}
{Szokoly}, G.~P., {et~al.} 2004, \apjs, 155, 271

\bibitem[{{Ulmer}(1999)}]{Ulmer99}
{Ulmer}, A. 1999, ApJ, 514, 180

\bibitem[{{Urry} \& {Padovani}(1995)}]{Urry95}
{Urry}, C.~M., \& {Padovani}, P. 1995, \pasp, 107, 803

\bibitem[{{Volonteri} {et~al.}(2005){Volonteri}, {Madau}, {Quataert}, \&
  {Rees}}]{Volonteri05}
{Volonteri}, M., {Madau}, P., {Quataert}, E., \& {Rees}, M.~J. 2005, \apj, 620,
  69

\bibitem[{{Wang} \& {Merritt}(2004)}]{Wang04}
{Wang}, J., \& {Merritt}, D. 2004, \apj, 600, 149

\bibitem[{{White} {et~al.}(1997){White}, {Becker}, {Helfand}, \&
  {Gregg}}]{FIRST97}
{White}, R.~L., {Becker}, R.~H., {Helfand}, D.~J., \& {Gregg}, M.~D. 1997,
  \apj, 475, 479

\bibitem[{{Wright}(2006)}]{Wright06}
{Wright}, E.~L. 2006, \pasp, 118, 1711

\bibitem[{{Yanny} {et~al.}(2009){Yanny}, {Rockosi}, {Newberg}, {Knapp},
  {Adelman-McCarthy}, {Alcorn}, {Allam}, {Allende Prieto}, {An}, {Anderson},
  {Anderson}, {Bailer-Jones}, {Bastian}, {Beers}, {Bell}, {Belokurov},
  {Bizyaev}, {Blythe}, {Bochanski}, {Boroski}, {Brinchmann}, {Brinkmann},
  {Brewington}, {Carey}, {Cudworth}, {Evans}, {Evans}, {Gates}, {G{\"a}nsicke},
  {Gillespie}, {Gilmore}, {Gomez-Moran}, {Grebel}, {Greenwell}, {Gunn},
  {Jordan}, {Jordan}, {Harding}, {Harris}, {Hendry}, {Holder}, {Ivans},
  {Ivezi{\v c}}, {Jester}, {Johnson}, {Kent}, {Kleinman}, {Kniazev},
  {Krzesinski}, {Kron}, {Kuropatkin}, {Lebedeva}, {Lee}, {Leger}, {L{\'e}pine},
  {Levine}, {Lin}, {Long}, {Loomis}, {Lupton}, {Malanushenko}, {Malanushenko},
  {Margon}, {Martinez-Delgado}, {McGehee}, {Monet}, {Morrison}, {Munn},
  {Neilsen}, {Nitta}, {Norris}, {Oravetz}, {Owen}, {Padmanabhan}, {Pan},
  {Peterson}, {Pier}, {Platson}, {Fiorentin}, {Richards}, {Rix}, {Schlegel},
  {Schneider}, {Schreiber}, {Schwope}, {Sibley}, {Simmons}, {Snedden}, {Smith},
  {Stark}, {Stauffer}, {Steinmetz}, {Stoughton}, {Subba Rao}, {Szalay},
  {Szkody}, {Thakar}, {Thirupathi}, {Tucker}, {Uomoto}, {Vanden Berk},
  {Vidrih}, {Wadadekar}, {Watters}, {Wilhelm}, {Wyse}, {Yarger}, \&
  {Zucker}}]{SEGUE09}
{Yanny}, B., {et~al.} 2009, \aj, 137, 4377

\bibitem[{{Zacharias} {et~al.}(2005){Zacharias}, {Monet}, {Levine}, {Urban},
  {Gaume}, \& {Wycoff}}]{NOMAD05}
{Zacharias}, N., {Monet}, D.~G., {Levine}, S.~E., {Urban}, S.~E., {Gaume}, R.,
  \& {Wycoff}, G.~L. 2005, VizieR Online Data Catalog, 1297, 0

\bibitem[{{Zhou} {et~al.}(2006){Zhou}, {Wang}, {Yuan}, {Lu}, {Dong}, {Wang}, \&
  {Lu}}]{Zhou06}
{Zhou}, H., {Wang}, T., {Yuan}, W., {Lu}, H., {Dong}, X., {Wang}, J., \& {Lu},
  Y. 2006, \apjs, 166, 128

\end{thebibliography}

\clearpage




\clearpage

\clearpage

\clearpage

\clearpage

\clearpage

\clearpage

\clearpage

\clearpage

\begin{table}
\begin{center}
\caption{X-ray Observations.\label{obs-1}}
\begin{tabular}{lrccccr}
\tableline\tableline

           &       &             &          & \multicolumn{2}{c}{Aim Point Coordinates} &  \\
\noalign{\vskip -6pt}
           &       & Observation & Duration & \multicolumn{2}{c}{\hrulefill} &  \\
Instrument & Obsid & Date (UT)   & (ks)     & $\alpha$(2000) & $\delta$(2000) & Roll Angle \\
\tableline

ACIS-I &540        & 2000 Apr. 15 & 10.5  & 13:11:13.88 & --01:20:14.75 &214.302  \\
ACIS-I &1663       & 2001 Jan. ~7 & 10.5  & 13:11:34.90 & --01:17:10.88 & 67.828  \\
EPIC   &0093030101 & 2001 Dec. 24 & 39.8  & 13:11:33.69 & --01:20:29.70 & 115.755 \\
ACIS-I &5004       & 2004 Feb. 28 & 19.9  & 13:11:33.30 & --01:17:20.91 & 75.154  \\
ACIS-I & 6930      & 2006 Mar. ~6 & 76.9  & 13:11:32.87 & --01:17:21.16 & 77.129  \\
ACIS-I &7289       & 2006 Mar. ~9 & 75.7  & 13:11:32.87 & --01:17:21.17 & 77.129  \\
ACIS-I &7701       & 2007 Mar. ~7 & ~5.1  & 13:11:37.04 & --01:19:13.76 & 77.454  \\

\tableline
\end{tabular}


\end{center}
\end{table}

\clearpage

\begin{sidewaystable}[!htpb]
\begin{center}
\caption{Variable Sources}\label{sigvar}
\begin{tabular}{ll@{}ll@{}cc@{}c@{}c@{}c@{}c@{}cl@{}c@{}c@{}l}

\tableline\tableline

Source\ \ & RA       & Dec       & SDSS Optical ID               & $I_{AB}$ &\multicolumn{6}{c}{\cha\ Obsid Coverage} & Peak Rate       & Peak HR\ \     & Peak $\theta$\ \ & Flags \\
 & (\degr)& (\degr) &                          & &540\ \ & 1663\ \ & 5004\  \ & 6930\ \ & 7289\ \ & 7701\ \ &  ($10^{-4}$ ct s$^{-1}$) & $\frac{H-S}{H+S}$ & (\arcmin)   &    \\
\tableline
14 & 197.9353     & -1.1661   & none                     & &     & X    & X    & P    & X    &      & $3.18\pm1.09$    & -0.38            & 12.8           & 1,2 \\
24 & 197.9825     & -1.1709   & J131155.51-011017.7 & 18.45 &     & P    & X    & X    & X    &      & $10.7\pm4.7$     & -0.39            & 8.68           & 4 \\
27 & 197.8700     & -1.1488   & J131128.78-010856.5 & 21.47 &     & X    & X    & X    & P    &      & $1.94\pm1.00$    & -0.38            & 8.49           & 1,3          \\
31 & 197.7475     & -1.1341   & J131059.20-010805.5 & 20.77 &     & X    & X    & X    & P    &      & $5.61\pm1.47$    & -0.52            & 12.5            & 1,2,4      \\
34 & 197.7035     & -1.1420   & J131049.00-010832.4 & 20.15 &     & X    & X    & X    & P    &      & $5.27\pm1.46$    & -0.42            & 14.1            & 1,2,3      \\
43 & 197.8888     & -1.2866   & J131133.28-011712.2 & 19.71 & P   & X    & X    & X    & X    & X    & $34.9\pm7.9$     & -0.69            & 5.74            & 2,3 \\
75 & 197.8232     & -1.2121   & J131117.54-011244.5 & 19.92 & X   &      & X    & P    & X    & X    & $0.26\pm1.07$    & { }1.00             & 6.00            & 4      \\
77 & 198.0102     & -1.3321   & none                     & &     & X    & P    & X    & X    & X    & $6.30\pm2.74$    & -0.04            & 7.73            & 1      \\
78 & 197.9978     & -1.2848   & J131159.55-011705.6 & 14.95 &     & P    & X    & X    & X    & X    & $11.7\pm5.1$     & -0.84            & 6.15            & 2,4      \\
86 & 197.9058     & -1.4036   & none                     & & X   & X    & X    & X    & P    & X    & $4.40\pm1.43$    & -0.52            & 6.95            & 1      \\
123 & 197.7654     & -1.4008  & none                     & & P   &      &      &      &      & X    & $15.0\pm5.6 $    & -0.61            & 4.58            & 2,6      \\
131 & 197.8080     & -1.4281  & none                     & & P   &      &      &      &      & X    & $16.9\pm5.9$     & -0.43            & 5.44            &       \\
134 & 197.8259     & -1.2416  & J131118.20-011429.8 & 18.36 & X   & X    & X    & X    & X    & P    & $277\pm26$       & -0.50            & 6.68            & 2,3,7,8      \\
{\bf 141} & {\bf 197.8424 }     & {\bf -1.3959 }  & {\bf J131122.15-012345.6 } & {\bf 19.60} & {\bf X}   & {\bf X}    & {\bf P}    & {\bf X}    & {\bf X}    & {\bf X}    & ${\bf 15.6\pm4.1}$     & {\bf -1.00}            & {\bf 6.98}            & {\bf 5}   \\
152 & 197.9071     & -1.3304  & J131137.68-011950.0 & 20.42 & P   & X    & X    & X    & X    & X    & $78.9\pm12$      & -0.64            & 5.97            & 2,4,8      \\
153 & 197.9144     & -1.3956  & J131139.42-012344.1 & 21.29 & P   & X    & X    & X    & X    & X    & $5.81\pm5.46$    & -1.00            & 7.29            & 2,4,7      \\
157 & 197.9352     & -1.3191  & J131144.39-011909.3 & 20.49 &     & X    & X    & P    & X    & X    & $20.1\pm20.0$    & -0.52            & 3.40            & 2      \\
160 & 197.9597     & -1.4248  & none                     &  &    & X    & P    & X    & X    & X    & $13.3\pm3.5$     & -0.28            & 9.19            & 2,6      \\
162 & 197.9695     & -1.2442  & none                     &  &    & X    & X    & P    & X    & X    & $9.16\pm1.52$    & -0.35            & 5.64            & 1,2,5      \\
163 & 197.9758     & -1.3797  & J131154.23-012246.9 & 19.99 & X   & P    & X    & X    & X    & X    & $23.3\pm6.0$     & -0.60            & 7.39            & 1,2,4,8      \\
166 & 197.9864     & -1.2198  & J131156.76-011311.8 & 13.26 &     & P    & X    & X    & X    & X    & $82.1\pm9.9$     & -0.92            & 6.76            & 2,4,8      \\
167 & 197.9928     & -1.1745  & J131158.32-011026.8 & 16.65 &     & X    & P    & X    & X    &      & $6.54\pm2.72$    &  { }0.62            & 9.29            & 1,5      \\
\tableline

\end{tabular}
\end{center}
{\small
Significantly variable sources and their SDSS optical counterparts, as described in \S2.2.1.  Bold indicates the flare described in detail in the text.  SDSS positions may differ slightly from X-ray source positions, but are within X-ray positional uncertainty.  $I_{AB}$ magnitudes are calculated from SDSS $ugriz$ magnitudes using \cite{Lupton05}.  Compare to $m^{\ast}=18.5$ from \cite{Lemze09} for $I_{AB}$.  X marks epochs when the object would be located within the ACIS FOV, regardless of whether it is detected during the epoch.  P indicates the epoch of peak count rate, for which the rate, HR and $\theta$ are presented.  H and S bands are as described in the text. \\

Flags:
(1) High pre-peak uncertainty
(2) XMM detected source
(3) SDSS flagged as possible QSO
(4) SDSS flagged likely star.  A small fraction may also be QSOs or other bright, unresolved objects.
(5) Possible pre-peak nondetection
(6) SDSS object visible but not catalogued
(7) Variability pattern inconsistent with flare
(8) Pre-existing match in 1RXH \citep{1RXH} or 2RXP \citep{2RXP}

}
\end{sidewaystable}

\clearpage




\end{document}